\documentclass[fleqn,usenatbib]{mnras}

\usepackage[T1]{fontenc}
\usepackage{ae,aecompl}

\usepackage{graphicx}	
\usepackage{amsmath}	
\usepackage{amssymb}	

\usepackage{newtxtext,newtxmath}

\title[The time delay of CLASS~B1600+434]{The time delay of CLASS~B1600+434 from VLA multi-frequency and polarization monitoring}

\author[A. D. Biggs]{
  A.~D.~Biggs$^{1}$\thanks{E--mail: abiggs@eso.org}
  \\
  $^{1}$European Southern Observatory, Karl Schwarzschild Stra{\ss}e 2, D-85748 Garching, Germany
}

\date{Accepted XXX. Received YYY; in original form ZZZ}

\pubyear{2021}

\begin{document}
\label{firstpage}
\pagerange{\pageref{firstpage}--\pageref{lastpage}}
\maketitle

\begin{abstract}
  We present an analysis of archival multi-frequency Very Large Array monitoring data of the two-image gravitational lens system CLASS~B1600+434, including the polarization properties at 8.5~GHz. From simulating radio light curves incorporating realistic external variability in image~A, we find time delays consistent at 1~$\sigma$ for all frequencies and in total flux density and polarization. The delay with the smallest uncertainty (total flux density at 8.5~GHz) is $42.3^{+2.0}_{-1.8}$ (random) $\pm 0.5$ (systematic)~d (equivalent to $42.3 \pm 2.1$~d) whereas combining all delay estimates gives a slightly higher value of $43.6\pm1.2$~d. Both values are lower than the previously published radio result and inconsistent with that found in the optical. $H_0$ determination is difficult due to the complicated lensing mass and the lack of constraints provided by only two images. However, analysis of archival Very Long Baseline Interferometry data reveals jets in this system for the first time, the orientations of which provide model constraints. In addition, extremely sensitive maps made from combining all the monitoring data reveal faint emission on one side of the lensing galaxy which we speculate might be the result of a naked-cusp lensing configuration. Finally, we find clear evidence for external variability in image~A on time-scales of days to years, the frequency-dependence of which supports the previous conclusion that this is predominantly due to microlensing. External variability seems to be completely absent in image~B and this does not appear to be a consequence of scatter-broadening in the interstellar medium of the lensing galaxy.
\end{abstract}




\section{Introduction}
\label{sec:intro}

\defcitealias{koopmans00a}{K00}

A measurement of the time delay between the images of a gravitational lens system potentially allows a one-step determination of $H_0$ using the technique first laid out by \citet{refsdal64}. In the radio, time delay measurements have been handicapped by a lack of intrinsic source variability and we have recently begun a reanalysis of archived Very Large Array (VLA) lens monitoring data with the goal of measuring the time delay via variability in polarization. Compared to total intensity, the magnitude of the polarization variability tends to be greater \citep[e.g.][]{saikia88} and a time delay can be measured using both the magnitude and position angle of the polarization. So far we have published a revised delay for JVAS~B0218+357 based on total and polarized flux density data \citep{biggs18} and the first time-delay measurement for JVAS~B1030+074 \citep{biggs18b} based on polarization data alone.

We now turn to the gravitational lens system B1600+434 which was discovered during the Cosmic Lens All-Sky Survey \citep[CLASS --][]{browne03} and consists of two images of a $z=1.589$ radio-loud quasar \citep{fassnacht98} with a separation of 1.4~arcsec and a flux ratio (A/B) of $\sim$1.2 \citep*{jackson95,koopmans98}. The main lensing galaxy is an edge-on spiral with a prominent dust lane \citep{jaunsen97,maller00} which forms part of a small group \citep{auger07}. Time delays have already been published for this system from monitoring data collected using the VLA at 8.5~GHz \citep[$47^{+5}_{-6}$~d --][hereafter \citetalias{koopmans00a}]{koopmans00a} and the Nordic Optical Telescope \citep[$51\pm2$~d --][]{burud00} -- 1-$\sigma$ uncertainties are quoted in each case. 

The VLA monitoring data originally analysed by \citetalias{koopmans00a} gave relatively poor constraints on the time delay as the only observed variability was a linear decline in the flux density. It was therefore necessary to combine these data with a measurement of the flux density ratio ($1.212 \pm 0.005$) taken from a previous season of VLA monitoring, the data from which remain unpublished. Also handicapping any attempt at measuring the time delay was the presence of additional variability in image~A. Variability seen in only one image cannot be intrinsic to the source and it is presumed that this external variability is due to a propagation effect which occurs as the radio waves pass through the $z=0.4144$ \citep{fassnacht98} lensing galaxy. \citet{koopmans00b} considered microlensing by compact halo objects the most likely explanation. The optical data are also subject to external variability, in this case microlensing by stars in the lensing galaxy \citep{burud00}.

B1600+434 was ultimately monitored at the VLA for a total of five observing seasons covering a period of just over six years. Multiple frequencies were often observed in order to investigate the nature of the external variability and together these data represent the most extensive lens monitoring campaign ever carried out in the radio. As the lensed source is also polarized in the radio \citep{patnaik01} these data provide an unprecedented number of constraints on the time delay in B1600+434. We have therefore undertaken a uniform reduction of all the archival VLA data and included, for the first time, an analysis of the polarization properties of the lensed source. An analysis of Seasons~2--5 of the 8.4-GHz data (total flux density only) appears in \citet{alicia09}.

In the next section we describe the available data and the different observing strategies used by the various observing teams. We then go on to detail the methods used to calibrate the various datasets before presenting the total flux density and polarization radio light curves. The methods used to derive the time delay and its associated uncertainty are explained in Section~\ref{sec:delay} and in Section~\ref{sec:discussion} we discuss the results, including the nature of the external variability. Finally, in Section~\ref{sec:lensmodel} we make use of archival \textit{Hubble Space Telescope} (\textit{HST}) and Very Long Baseline Interferometry (VLBI) data to investigate possible improvements to the lens model, before presenting our conclusions in Section~\ref{sec:conclusions}.

\begin{table*}
  \centering
  \caption{Summary of the five VLA monitoring seasons of B1600+434. The number of epochs refers to observations at 8.5~GHz -- other bands can differ slightly. $^{\dagger}$~In C-configuration, the angular resolution at 5 and 8.5~GHz was not sufficient to derive reliable flux densities for both images separately. $^{\ddagger}$~Two epochs of AX004 (15 and 16 March) were taken as part of program AM593, the third observing run of CLASS. The first was not usable due to snow. An additional epoch (8 May) is available in the archive under the code ADA000.}
  \label{tab:obs}
  \begin{tabular}{cccccc} \\ \hline
    Season & Project code & Dates & Configuration(s) & Frequency band(s) (GHz) & Number of epochs \\ \hline
    1 & AH593 & 1996 Oct 10 to 1997 May 24 & A, B & 8.5, 15 & 44 \\
    2 & AX004$^{\ddagger}$ & 1998 Feb 13 to 1998 Jun 09 & A & 8.5 & 40 \\
      & AF340 & 1998 Jun 11 to 1998 Oct 19 & B & 8.5 & 39 \\
    3 & AB922 & 1999 Jun 15 to 2000 Feb 14 & A, B & 1.4, 5, 8.5 & 90 \\
      & AK507 & 2000 Feb 19 to 2000 Jun 20 & C & 8.5$^\dagger$, 15, 22 & 39 \\
    4 & AK518 & 2000 Oct 06 to 2001 May 29 & A, B & 1.4, 5, 8.5, 15 & 66 \\
    5 & AK543 & 2002 Jan 17 to 2002 Sep 28 & A, B & 1.4, 5, 8.5, 15 & 78 \\
      & AK554 & 2002 Oct 01 to 2002 Dec 28 & C & 5$^\dagger$, 8.5$^\dagger$, 15 & 32 \\ \hline
  \end{tabular}
\end{table*}

\section{Observations and data reduction}
\label{sec:obs}

In all, eight separate proposals have been accepted by the VLA to monitor B1600+434 (Table~\ref{tab:obs}). The first was AH593, a project which monitored eight lens systems and which led to the publication of a time delay for JVAS~B0218+357 \citep{cohen00}. Although never published, the 1600+434 data allowed the measurement of a flux density ratio which was subsequently used by \citetalias{koopmans00a} to determine the time delay from their own VLA monitoring data. All data apart from the first season's were collected by the Koopmans group and a preliminary analysis of the 5-GHz AB922 data was presented by \citet{koopmans01}. A total of 428 epochs were observed.

One aspect of the scheduling that is relevant to the time-delay estimation is that consecutive epochs were generally separated by very close to an integer multiple of 1~d. The most common separation between epochs was 3~d but if the separations are placed in bins with a width of 0.1~d there are clear peaks at 1.0, 2.0, 3.0, 4.0, etc.~d. As described in Section~\ref{sec:delay}, this leads to features in the $\chi^2$ and dispersion minimization spectra that can lead to quantization of the best-fit delays.

In all cases, the standard VLA continuum setup was used comprising two 50-MHz-wide subbands recording left and right circular polarizations (RCP and LCP) and the correlator produced all four correlations necessary for polarization analysis. Only 8.5~GHz was observed at every epoch. The AH593 monitoring also included observations at 15~GHz whilst the Koopmans group added 1.4, 5, 15 and 22~GHz (the latter during one season only) once the presence of the external variability became apparent. Observations were conducted in the VLA's A, B and C configurations, although in the latter it is not possible to reliably determine the flux densities of A and B separately for $\nu \le 8.5$~GHz due to the poor angular resolution.

\subsection{Observing Strategies}

The observing strategies of the two groups (AH593 and Koopmans) differed in various aspects. AH593 was based on the use of a `traditional' calibration approach including scans on a standard flux calibrator (3C~286), a bright unpolarized source for determination of the instrumental polarization (OQ~208) and a phase calibrator (J1625+4134, also known as 4C~41.32). 3C~286 was also used for calibration of the absolute angle of the electric vector position angle (EVPA).

The Koopmans observations are more difficult to describe as the observing strategy evolved with time and was not the same at each frequency. One dominant feature was the near-exclusive use of J1634+6245 (3C~343) for flux calibration. This is a compact steep-spectrum source for which VLBI observations reveal a peculiar structure consisting of several compact components and a possible jet embedded in a diffuse structure with a total extent of $\sim$100~mas \citep{mantovani10}. Its flux density was shown to be non-variable in an earlier lens monitoring programme \citep{fassnacht99b} at 8.5~GHz. As many of the AB922 and AK507 epochs include 3C~286 we can directly measure the 3C~343 flux density and find that this actually slowly increases by $\sim$2.5~per~cent over a period of a year (1999 June -- 2000 June) with an average flux density of $\sim$0.785~Jy. Our own reduction of the last few epochs of the \citet{fassnacht99b} data (May 1997) give a very similar value (0.78--0.79~Jy). We therefore conclude that this source indeed varies only slowly by a few per~cent per year in total flux density, a conclusion supported by multi-frequency monitoring with the University of Michigan Radio Astronomy Observatory \citep*[UMRAO --][]{aller92}\footnote{We note, however, that \citetalias{koopmans00a} used a higher 8.5-GHz total flux density of 0.84~Jy measured with the Westerbork Synthesis Radio Telescope (WSRT) in December 1998.}. The epochs including 3C~286 have also allowed us to calculate the flux density of this source in the other observing bands.

3C~343 was not observed at 1.4~GHz during the AB922 monitoring and at this frequency the most suitable flux calibrator that is always present is J1559+4349, a steep-spectrum double with a weak core \citep{yan16}. Its flux density was measured from those AB922 epochs which included 3C~286 and found to be non-time-variable. This source was also often observed at other frequencies where it could be used as a secondary flux calibrator i.e.\ to check on the quality of the 3C~343-based flux calibration. Its close proximity to B1600+434 ($\Delta\theta<$1\degr) makes it particularly effective in this role. Another secondary calibrator was J1545+4751, a compact core-jet source \citep{tremblay16} which appears to not vary in total flux density. It was predominantly observed at 15~GHz and is 5.4\degr from B1600+434.

A number of sources suitable for calibration of the instrumental polarization were included during the Koopmans monitoring. The most useful was the Compact Symmetric Object (CSO) J1400+6210 which was observed during nearly every epoch. Many AX004 epochs did not include this source and so another CSO (J1944+5448) was used instead. CSOs make good instrumental polarization calibrators as they are very weakly polarized \citep{peck00}. Unfortunately, no recognised EVPA calibrator was regularly observed but we have determined that 3C~343 functions acceptably in this role. Although not greatly polarized, $\sim$1~per~cent, its polarized flux density ($\sim$10~mJy at 8.5~GHz) results in a signal-to-noise ratio (SNR) between 3 and 5 times higher than that of B1600+434, depending on the observing band. 3C~343 is stable in polarized flux density and EVPA, the latter varying at 8.5~GHz by only $\sim$4\degr over the nine-month duration of AB922. These data could be calibrated using 3C~286 and we measure an average EVPA of 68, 68 and 42\degr at 5, 8.5 and 15~GHz respectively. At 1.4~GHz 3C~343 appears to be rather variable in polarization and thus we have not calibrated the polarization properties at this frequency. Observations of B1600+434 were always bracketed by scans on an unresolved phase calibrator.

\subsection{Calibration}

All data were calibrated using NRAO's Astronomical Image Processing System (\textsc{aips}). We first set the total flux density of the flux calibrator and then derived complex gain solutions for all calibrators, a clean-component model being used if any source was significantly resolved. Bad data were often found at this point and the gain calibration repeated once these were flagged. The flux density of all calibrators was then found relative to the flux calibrator using \textsc{getjy}. At this point it was usually the case that the gain solutions were transferred from the unresolved phase calibrator to B1600+434. However, at 15~GHz it was found that more accurate flux calibration was possible by fitting a smooth function (usually linear) to the phase calibrator flux densities and then repeating the gain calibration using the updated flux densities. Another exception was 1.4~GHz where the gain solutions were transferred directly from the flux calibrator to the lens as the separation between these sources is very small ($<$1\degr).

With the flux calibration completed, the antenna-based instrumental polarization `D-terms' were then calibrated using \textsc{pcal}. Polarization maps were made of 3C~343 at each epoch and frequency and the difference between the measured and expected EVPA used to set the absolute angle of polarization of each epoch's data (RCP $-$ LCP phase difference) using \textsc{rldif}.

\subsection{Modelfitting}

The Stokes $I$, $Q$ and $U$ flux densities of the calibrated B1600+434 data were calculated using the Caltech \textsc{difmap} program \citep{shepherd97}. Delta components representing the two lensed images were fitted to the $u,v$ data with their positions fixed and only the fluxes allowed to vary. Modelfitting was first done to the Stokes $I$ data with multiple iterations of modelfitting and phase self-calibration. The Stokes $Q$ and $U$ flux densities were then fitted to the self-calibrated data. At 1.4~GHz it was necessary to include other sources in the model due to the large primary beam at this frequency. This was done by making a faceted wide-field map from a multi-epoch dataset in \textsc{aips} and then subtracting all clean components other than those associated with the lens from each individual 1.4-GHz epoch using \textsc{uvsub}. The resultant modelfitted B1600+434 flux densities are significantly smoother once this step is included.

Flux density errors were calculated from a quadrature addition of the noise in a naturally weighted residual map (with the two lensed images subtracted) and an overall flux-scale error. This was set to 0.5~per~cent (1~per~cent at 15~GHz) based on the scatter in the calibrated flux densities of sources that should either vary slowly or not at all with time, particularly the secondary flux calibrators J1559+4349 and J1545+4751. Polarization errors were calculated as detailed in \citet{biggs18b} and include a contribution from residual leakage (0.1~per~cent of the total flux density) with error combination and correction for Rician bias (the magnitude of polarization is positively biased at low SNR) performed following the formulation of \citet{wardle74}. For one epoch of image~A the polarization flux density was less than its error which makes the correction for Rician bias undefined -- this measurement was therefore excluded from the time-delay analysis.

The EVPA error contains a term added in quadrature equal to the uncertainty in the EVPA of the calibrator. This was calculated from the SNR in the polarization maps and rarely exceeds 1\degr\ for 3C~343. The errors are much smaller for the much brighter calibrators used for the AH593 data but we enforce a minimum error of 0.5\degr\ based on our experience with B0218+357 \citep{biggs18}. A final correction was made to the AH593 EVPA values related to the proximity of the calibrator (3C~286 for all but one epoch) to the zenith at transit. Again, for more details see \citet{biggs18}.

Unreliable epochs were identified and removed from the variability curves. These usually have a poor fit to the $u,v$ data (as measured by $\chi^2$) or higher than usual noise in the residual map, both of which are usually attributable to bad weather as indicated in the observing logs. In some cases, only the polarization measurements were deleted because e.g.\ no D-term calibrator was present during that epoch.

During the AB922 monitoring both secondary flux calibrators (J1559+4349 and J1545+4751) were observed at 8.5~GHz. An inspection of their flux densities (both derived from 3C~343) showed that they were correlated with the same variability pattern also being present in the B1600+434 data. Therefore, as a final step we corrected these B1600+434 flux densities using the weighted average of the normalised flux densities of the two secondary calibrators.

\section{Variability curves}
\label{sec:results}

\begin{figure*}
\begin{center}
\includegraphics[width=\linewidth]{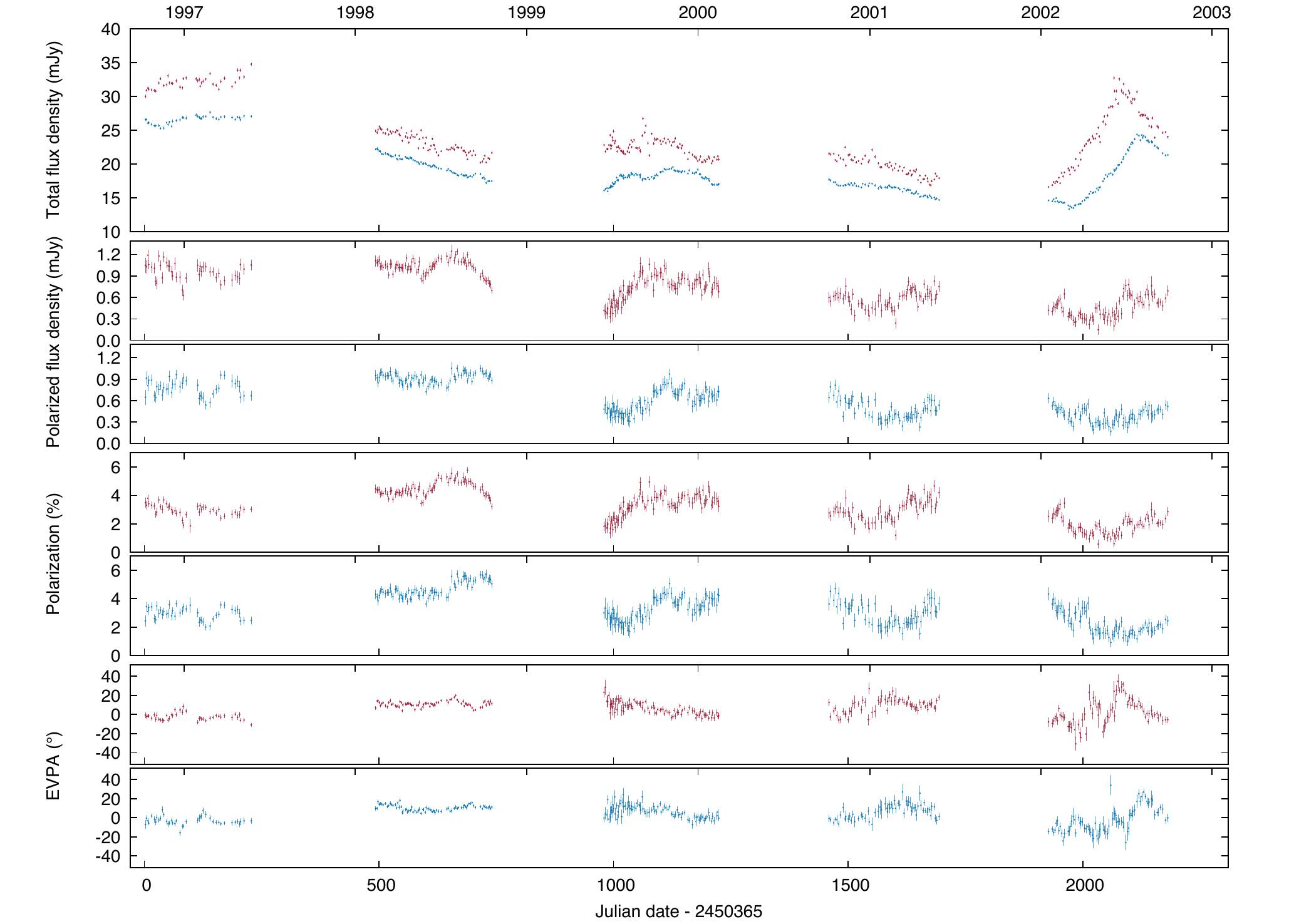}
\caption{VLA 8.5-GHz variability curves of B1600+434 for each of the five seasons of monitoring. From top to bottom: total flux density, polarized flux density, percentage polarization and EVPA. For clarity, images A (red symbols) and B (blue symbols) are shown in separate panels for the polarization plots.}
\label{fig:vc_x}
\end{center}
\end{figure*}

Of the four frequency bands observed during more than one season, only 8.5~GHz was continuously present and the variability curves at this frequency are shown in Fig.~\ref{fig:vc_x}. The flux density of image~B varies very smoothly and short-term variability (e.g.\ over time-scales $\la$10~days) is almost completely absent. Epochs occasionally deviate from the smooth variation and we believe that this is most likely due to measurement error.

Image~A, by way of contrast, appears much noisier than B due to the external variability that was originally reported by \citet{koopmans00b}. This can be seen to be present throughout the monitoring -- at no point is the variability of image~A as smooth as that of image~B. The external variability is though time-dependent and the `caustic-crossing event' towards the end of 1999 \citep{koopmans01} is a notable example of where larger variations occur over a relatively short time period ($\sim$30~days).

\begin{figure*}
\begin{center}
\includegraphics[width=\linewidth]{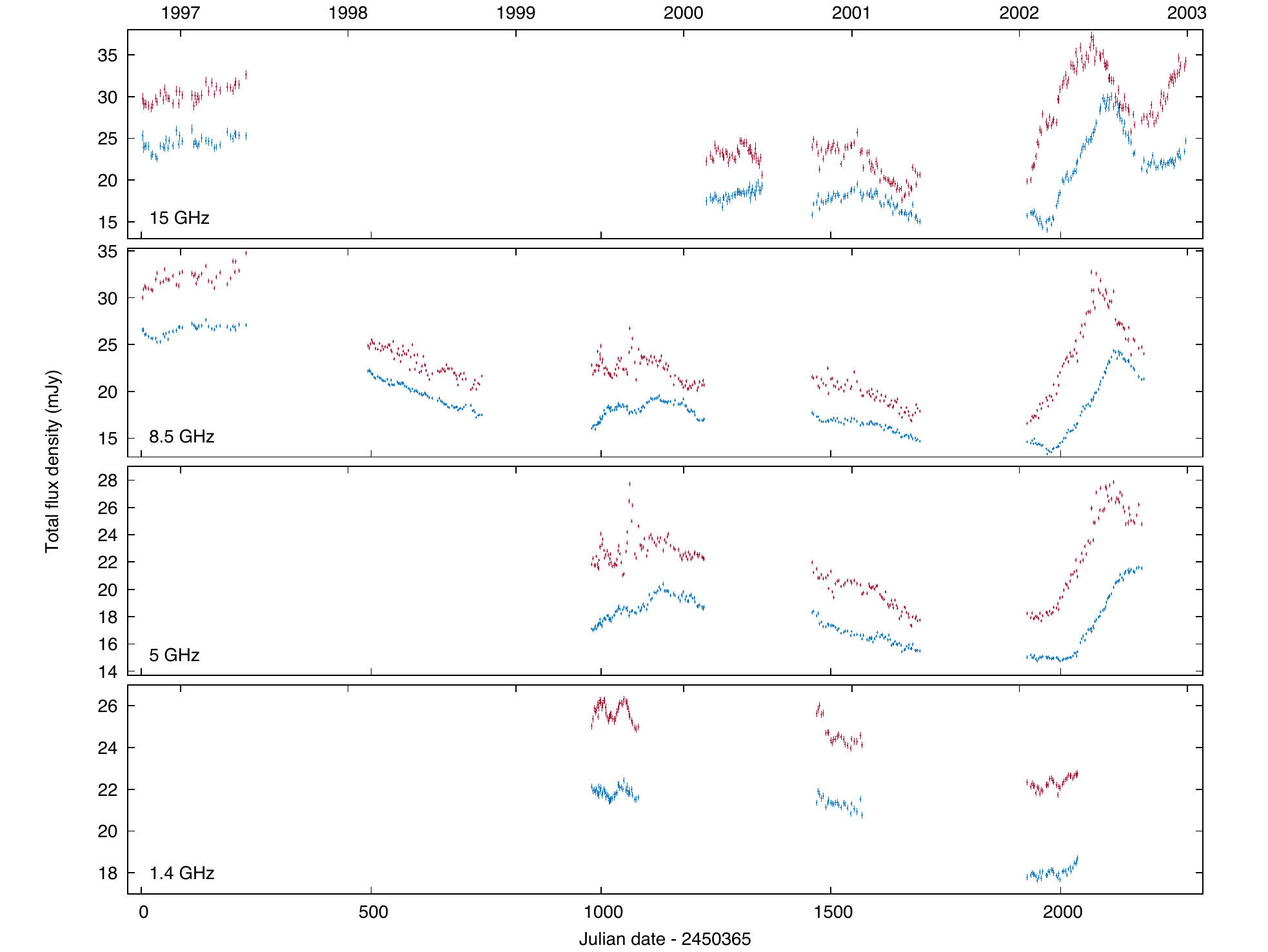}
\caption{VLA multi-frequency variability curves of B1600+434 in total flux density. From top to bottom: 15, 8.5, 5 and 1.4~GHz. Image~A (red symbols) is the brighter image in each case. There is little 1.4-GHz data because the low frequency requires the largest VLA configuration (A) in order to cleanly resolve the lensed images. At 15~GHz it is possible to monitor in the much smaller C configuration and thus the last season is particularly long at this frequency.}
\label{fig:vc_all_tf}
\end{center}
\end{figure*}

We show the multi-frequency total flux density data in Fig.~\ref{fig:vc_all_tf} where it can be seen that the external variability (the extra `noise' in image~A) is also present at all frequencies. A hint of a frequency dependence to the external variability is provided by the `caustic-crossing event'. This is very obvious at both 8.5 and 5~GHz, but is not immediately apparent at 1.4~GHz. It is possible that this corresponds to the second, rather smooth, peak in this season's 1.4-GHz data, but the lack of later data in image~B make this impossible to verify. However, a clear example of external variability \textit{is} seen at 1.4~GHz during Season~4 -- a very prominent drop in flux density is seen in image~A shortly after the beginning of Season~4 that appears to lack a counterpart in B 40--50~days later.

The most prominent feature of the \textit{intrinsic} source variability is the rapid and large increase in the source brightness during Season~5. This is greatest at 15~GHz where the flux density doubles over a period of $\sim$130~days. The magnitude of the event decreases towards lower frequencies and also begins later, both characteristics of a flare caused by a relativistic shock in a conical jet \citep[e.g.][]{marscher85} -- see also the synthetic multi-frequency radio light curves of \citet*{fromm15}. At 1.4~GHz it is not clear that any part of the flare is detected before the monitoring at this frequency ceases. As well as showing the greatest variability, the 15~GHz data also continue for longer and allow the detection of a further phase of brightening.

Presented for the first time are the 1996--1997 data (Season~1) where the low level of variability was assumed to allow a robust measurement of the flux density ratio at 8.5~GHz. This was the key piece of information that allowed \citetalias{koopmans00a} to determine the time delay from the second season's data  and indeed the lensed source was relatively quiescent during this period, although systematic variations are present.

The 8.5-GHz polarization data are also shown in Fig.~\ref{fig:vc_x}. With a maximum polarization of $\sim$6~per~cent, the SNR of the polarization data is much lower than that of the total flux density, but this is compensated for by the greater variability. The polarized flux density varies from the lowest detectable value of $\sim$0.2~mJy to more than a factor of six higher. The EVPA variations are generally smoother, but in the last season multiple oscillations are seen which increase in magnitude as the flare in total flux density reaches its peak -- swings of 50\degr are seen with a typical time-scale of $\la$40~days.

The power of polarization monitoring for time-delay determination is illustrated by Season~2. These are the \citetalias{koopmans00a} data which alone were unable to constrain the time delay due to the linear nature of the total flux density variations. The polarization data, however, very obviously constrain the time delay due to a rapid change in the magnitude of polarization approximately halfway through this season's monitoring that is visible in both images -- the polarization increases from 3.5 to 5.5~per~cent in about one month with a typical measurement error of 0.25~per~cent. The first season also includes multiple oscillations in both polarization magnitude and EVPA whilst being relatively stable in total flux density.

At the same time, the polarization data show no obvious signs of the external variability that is so prominent in the total flux density of image~A and which makes it more difficult to determine the time delay. However, this is mainly a consequence of the low SNR of the polarization data and we shall show that the polarized flux density in particular is affected by significant levels of external variability in image~A (Section~\ref{sec:external}). We do not show nor consider further the polarization data in the other frequency bands due to a combination of low SNR and lower levels of variability.

In summary, although the lensed source shows little short-term variability in total flux density, on long time-scales the variability is significant and clearly sufficient to allow the measurement of a time delay. This is particularly true of the flare in Season~5 although at 5~GHz it is unfortunate that the monitoring ended before the flux density began to decline again in image~B. The usefulness of the 1.4-GHz data for time-delay determination is not clear although the Season~3 data seem to provide genuine constraints. The flare in Season~5 was also accompanied by significant variability in the EVPA and the polarization data generally show greater variability than total flux density.

\section{Time-delay analysis}
\label{sec:delay}

\subsection{Methods}

To measure the time delay from the presented data we have used two techniques: chi-squared minimization \citep[CSM -- e.g.][]{press92,kundic97,biggs99,fohlmeister07} and the dispersion minimization method of \citet{pelt94,pelt96}\footnote{A note on nomenclature: the `dispersion minimization' technique of \citet*{tewes13} which was first used in \citet{courbin11} differs significantly from Pelt's algorithm in that it removes the non-parametric basis of the method and instead uses interpolation. The Tewes method is actually identical to the well-established chi-squared minimization technique.}. The main difference in these approaches is whether interpolation is used or not.

The CSM method shifts the data of one image by a trial time delay and uses linear interpolation to create and compare a matching value in the other image. For a given delay $\tau$ and $N$ epochs in image A with interpolated equivalents in B, for e.g.\ flux density data we calculate the reduced chi-squared ($\chi^2$ per degree of freedom or $\chi^2_{\mathrm{dof}}$) as
\begin{equation}
  \label{eq:csm}
  \chi^2_{\mathrm{dof}} = \frac{1}{N_{\mathrm{dof}}} \sum_i^N \frac{\left(f_{\mathrm{A},i}(t+\tau) - r \tilde{f}_{\mathrm{B},i}(t)\right)^2}{\sigma^2_{\mathrm{A},i} + r \tilde{\sigma}^2_{\mathrm{B},i}}
\end{equation}
where $r$ is the A/B flux ratio, $\tilde{f}$ indicates an interpolated measurement and $_{\mathrm{dof}}$ is the number of degrees of freedom. $\chi^2_{\mathrm{dof}}$ is also calculated by shifting B by $-\tau$ and interpolating in image A -- the two $\chi^2$ values are then averaged. Note that no interpolation is done in the gaps between monitoring seasons -- if any points are shifted into a gap they are ignored and do not contribute to the goodness-of-fit statistic. $N$ and $N_{\mathrm{dof}}$ are in general functions of the delay as the number of points that can be interpolated changes with the amount of overlap between the shifted and unshifted time series. For EVPA data the equivalent calculation is
\begin{equation}
\chi^2_{\mathrm{dof}} = \frac{1}{N_{\mathrm{dof}}} \sum_i^N \frac{\left(f_{\mathrm{A},i}(t+\tau) - \left(\tilde{f}_{\mathrm{B},i}(t) + \Delta \theta\right)\right)^2}{\sigma^2_{\mathrm{A},i} + \tilde{\sigma}^2_{\mathrm{B},i}}
\end{equation}
i.e.\ instead of scaling the image~B value, the $y$-offset is removed by adding a constant offset ($\Delta \theta$) to the image~B EVPA value -- the associated error is not changed. The EVPA offset might represent differential Faraday rotation \citep[as is the case in B0218+357 --][]{patnaik93,biggs18} or differential magnification across multiple components of a source that have different intrinsic EVPAs.

The best-fit delay is found using a brute-force approach i.e.\ calculating $\chi^2_{\mathrm{dof}}$ for a grid of trial delays between 30 and 60~d with a sampling interval of 0.1~d. For each delay the magnification is solved for using the Powell method of optimization available in \textsc{scipy}. The delay with the lowest $\chi^2_{\mathrm{dof}}$ is the most likely delay.

The Pelt method offers an alternative approach as it does not use interpolation and instead compares actual measurements from different images. All epochs of image~A are first shifted by the trial delay and those from image~B scaled or rotated by the $y$-offset. The A and B measurements are then formed into a combined light curve $C$ containing $N$ points ordered by increasing time. Of the various algorithms provided by \citet{pelt96} we have used both the $D^2_3$ and $D^2_{4,2}$ statistics. The former calculates the dispersion by comparing neighbouring pairs as
\begin{equation}
D^2_3 = \frac{\sum_{k=1}^{N-1} S_k W_k G_k \left(C_{k+1} - C_k\right)^2}{2 \sum_{k=1}^{N-1} S_k W_k G_k}
\end{equation}
where $W_{n,m}$ gives the statistical weight of a pair ($1/(\sigma_n^2 + \sigma_m^2)$) and $G_{n,m} = 1$ if the two points in a pair are from different images and 0 otherwise. The term $S_k = 1$ if points are within some minimum distance of each other and 0 otherwise. We used this term only to prevent comparison of points lying on opposite sides of a season boundary.

The second statistic is more elaborate and the one most often used in the literature. It is similar to the first but calculates the weighted average of the squared difference between all A, B pairs as
\begin{equation}
D^2_{4,2} = \frac{\sum_{n=1}^{N-1} \sum_{m=n+1}^{N} S_{n,m} W_{n,m} G_{n,m} \left(C_n - C_m\right)^2}{\sum_{n=1}^{N-1} \sum_{m=n+1}^{N} S_{n,m} W_{n,m} G_{n,m}}.
\end{equation}
The $S$ weighting term is now a smooth function of the time separation between the two points in each pair,
\begin{equation}
S_{n,m} = \frac{1}{1 - \frac{| t_{\mathrm{n}} - t_{\mathrm{m}} |}{\delta}},
\end{equation}
and pairs which are separated by more than the free parameter $\delta$ do not contribute to the dispersion at all i.e.\ $S_{n,m} = 0$. The other terms have the same meaning as before.

The spectrum of the $D^2_{4,2}$ statistic (dispersion as a function of delay) is smoother than that produced by $D^2_3$ and has fewer local minima, but requires a decision as to which value of $\delta$ to use. For both Pelt methods, the best-fit delay itself was solved by iteratively optimizing the delay (using a simulated-annealing algorithm, `dual annealing', provided by \textsc{scipy}) and the $y$-offset parameters (using the same Powell minimization method used with the CSM method).

\subsection{Results for individual seasons}

\begin{table*}
  \centering
  \caption{Results of the two time-delay analysis techniques: chi-squared minimisation (CSM) and the Pelt dispersion algorithm including only neighbouring pairs ($D_3^2$). For the individual seasons, best-fit time delays are found throughout the trial delay range of 30--60~d reflecting the sometimes poor constraints resulting from either low intrinsic variability or high external variability in image A. The latter is responsible for the high $\chi^2_{\mathrm{dof}}$. The analysis of each season's data shows that the total flux density ratio (A/B) increases with time (see also Fig.~\ref{fig:fluxr}). No such effect is visible in the polarization data. Delays found by applying the CSM and Pelt techniques to the data from all available seasons simultaneously are shown in bold. No $y$-offset is shown for the combined total flux density data as this is modelled as a first-order polynomial. TF: total flux density, PF: polarized flux density.}
  \label{tab:delay}
  \begin{tabular}{ccc@{\hskip 1.5cm}ccc@{\hskip 1.5cm}cc} \\ \hline
    Data & Season & $\nu$ (GHz) & \multicolumn{3}{c}{CSM} & \multicolumn{2}{c}{$D^2_3$} \\
            &        &             & Delay (d) & $y$-offset & $\chi^2_{\mathrm{dof}}$ & Delay (d) & $y$-offset \\ \hline
    TF     & 3   & 1.4 & 45.5 & 1.174 &  0.9 & 46.8 & 1.176 \\
           & 4   & 1.4 & 36.9 & 1.175 &  5.5 & 36.6 & 1.174 \\
           & 5   & 1.4 & 52.7 & 1.225 &  1.3 & 51.0 & 1.223 \\
           & \textbf{all} & \textbf{1.4} & \textbf{48.7} & \textbf{$-$} & \textbf{2.8} & \textbf{46.8} & \textbf{$-$} \\
           & 3   & 5   & 51.4 & 1.207 & 16.0 & 51.5 & 1.208 \\
           & 4   & 5   & 59.3 & 1.243 &  4.3 & 56.4 & 1.238 \\
           & 5   & 5   & 41.4 & 1.234 &  7.9 & 42.2 & 1.232 \\
           & \textbf{all} & \textbf{5} & \textbf{44.6} & \textbf{$-$} & \textbf{10.9} & \textbf{45.3} & \textbf{$-$} \\
           & 1   & 8.5 & 45.6 & 1.191 &  2.2 & 41.2 & 1.194 \\
           & 2   & 8.5 & 40.7 & 1.207 &  4.1 & 41.1 & 1.206 \\
           & 3   & 8.5 & 46.4 & 1.240 &  6.7 & 46.5 & 1.240 \\
           & 4   & 8.5 & 45.4 & 1.247 &  3.2 & 45.1 & 1.242 \\
           & 5   & 8.5 & 42.3 & 1.264 &  6.4 & 40.8 & 1.264 \\
           & \textbf{all} & \textbf{8.5} & \textbf{42.3} & \textbf{$-$} & \textbf{5.3} & \textbf{42.8} & \textbf{$-$} \\
           & 1   & 15  & 34.4 & 1.218 &  0.9 & 37.8 & 1.219 \\
           & 3   & 15  & 46.7 & 1.246 &  0.7 & 46.0 & 1.248 \\
           & 4   & 15  & 44.1 & 1.284 &  2.1 & 44.7 & 1.280 \\
           & 5   & 15  & 48.7 & 1.272 &  4.2 & 48.1 & 1.271 \\
           & \textbf{all} & \textbf{15} & \textbf{44.5} & \textbf{$-$} & \textbf{2.8} & \textbf{46.7} & \textbf{$-$} \\ \hline
    PF     & 1   & 8.5 & 48.2 & 1.199 & 1.0 & 49.0 & 1.205 \\
           & 2   & 8.5 & 39.4 & 1.167 & 0.8 & 40.1 & 1.163 \\
           & 3   & 8.5 & 42.9 & 1.189 & 0.7 & 41.0 & 1.167 \\
           & 4   & 8.5 & 52.2 & 1.232 & 0.9 & 51.8 & 1.188 \\
           & 5   & 8.5 & 43.3 & 1.198 & 0.8 & 45.6 & 1.116 \\
           & \textbf{all} & \textbf{8.5} & \textbf{40.5} & \textbf{$-$} & \textbf{1.1} & \textbf{46.9} & \textbf{$-$} \\ \hline
    EVPA   & 1   & 8.5 & 41.5 & 2.46\degr  & 0.5 & 44.7 & 2.25\degr \\
           & 2   & 8.5 & 48.1 & 1.96\degr  & 1.2 & 47.9 & 1.61\degr \\
           & 3   & 8.5 & 50.2 & 1.96\degr  & 0.7 & 50.2 & 1.85\degr \\
           & 4   & 8.5 & 38.4 & 2.00\degr  & 1.1 & 31.8 & 2.26\degr \\
           & 5   & 8.5 & 44.4 & 2.51\degr  & 0.8 & 44.5 & 2.72\degr \\
           & \textbf{all} & \textbf{8.5} & \textbf{45.5} & \textbf{$-$} & \textbf{1.2} & \textbf{44.9} & \textbf{$-$} \\ \hline
  \end{tabular}
\end{table*}

The results from the two different techniques are shown in Table~\ref{tab:delay}. Best-fit time delays are found that cover the entire range of trial delays, testifying to the fact that many of the datasets are individually unable to place strong constraints on the delay. In general both techniques give similar results with delays that differ by only a few days, with a maximum difference of 6.6~d for the fourth season of the 8.4-GHz EVPA. The 8.5-GHz total flux density data of \citetalias{koopmans00a} (Season~2) give very consistent results (40.7 and 41.1~d) and do actually contain enough intrinsic variability to render an independent measurement of the total flux density ratio unnecessary.

The effect of the external variability in image~A is clearly reflected in the high values of reduced $\chi^2$. This statistic should be equal to unity for the case of two variability curves that show good agreement within the errors, but for the total flux density data this is almost never the case. The highest $\chi^2$ is the value of 16.0 seen during Season~3 at 5~GHz which corresponds to the `caustic-crossing event' \citep{koopmans01} but Season~5 also has consistently high values of reduced $\chi^2$ at all frequencies. The polarization data, on the other hand, all have $\chi^2 \sim 1$ which is to be expected given the lack of any obvious signs of external variability in these data.

Ultimately, the best constraints should be obtained when running the time-delay analysis on the data from all seasons simultaneously. The results from the individual seasons could in principle be combined using some form of weighted average, but this is likely to give a poorer estimate of the delay as an individual season is more prone to bias due to the effects of the external variability and the often limited intrinsic variability -- see for example the 5-GHz total flux density results for Seasons~3 and 4. As we shall see, there are also systematic effects related to the epochal sampling which are likely correlated between different seasons and which would complicate any averaging process.

However, in finding the delay using all seasons simultaneously it is necessary to account for any possible change in the $y$-offset between A and B that might occur over such a long period of monitoring. In fact, a clear trend is seen in the total flux density ratio, particularly at 8.5~GHz where this increases with each successive season. A similar trend appears to hold at the other frequencies, but interpretation is potentially made more difficult by the strong degeneracy between delay and flux ratio -- as the delays from the individual seasons show a large scatter, there will also be increased scatter in the flux density ratios.

\begin{figure}
\begin{center}
\includegraphics[width=\linewidth]{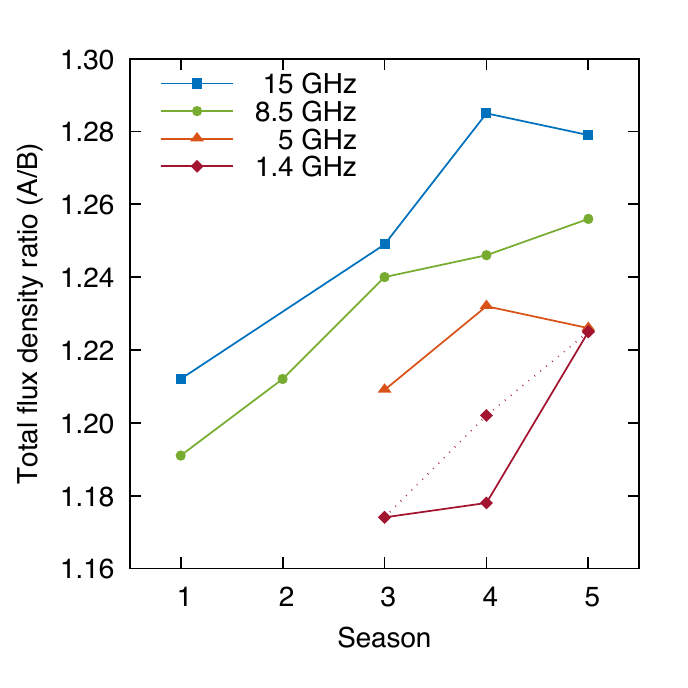}
\caption{Flux density ratio (A/B) derived from the CSM analysis for each season of the total flux density data at 1.4, 5, 8.5 and 15~GHz. For consistency, all values correspond to the same time delay. The flux ratio increases with both frequency and time and at 8.4~GHz this is very nearly linear. Discerning the true long-term trend in the flux ratio is made more difficult by the presence of short-term external variability and at 1.4~GHz we also show a point calculated using only the early epochs of Season~4 -- the resulting trend (dotted line) is much more linear.}
\label{fig:fluxr}
\end{center}
\end{figure}

To give a consistent view of any trend in the flux density ratio, in Fig.~\ref{fig:fluxr} we show this as measured using the CSM method for all the individual datasets but measured at the same delay (45~d, the centre of the delay search range). The 8.5-GHz data provide the clearest view of this effect as there are data in all five seasons and it can be seen that the flux density ratio is increasing approximately linearly with time. The flux ratio in Season~3 is too high compared to the linear trend, but this is unsurprising given the large magnitude of the short-time-scale external variability during this season which predominantly increases the flux density of image~A. Similarly, the sudden drop in the flux density of image~A during Season~4 at 1.4~GHz makes the flux density ratio here very uncertain. If we assume that the early A epochs are unaffected by the short-time-scale external variability then the flux density ratio for this season would be approximately 2-per-cent higher (shown as a corrected trend in Fig.~\ref{fig:fluxr}).

Given that we already know that external variability is causing the A/B flux density ratio to vary with time on short time-scales, the most likely explanation for this change on a time-scale of years is that it is caused by the same process.

\subsection{Multi-season results}
\label{sec:mseasonres}

The observed slow variation in the flux density ratio can be accounted for in a number of ways, for example by modelling the additional variability with a spline or polynomial \citep[e.g.][]{lehar92,kochanek06,tewes13}. We have removed the long-term linear variation of the total flux density ratio by modelling this as a linear function of time for both the Pelt and CSM methods. No obvious trend with time is seen with the polarized flux density ratio or the EVPA offset and therefore we did not fit a smooth function to represent any change in these quantities. Instead, we optimized a single value of the ratio or offset per season, an approach previously followed by e.g.\ \citet{fassnacht02}.

The resulting multi-season delays are shown as bold text in Table~\ref{tab:delay} and show much more consistency (40.5--48.7~d) than the single-season values. This is as expected given that all features common to the A and B variability curves are now contributing to the minimisation statistic simultaneously and act together to reduce the effect of the external variability. This prevents unrelated features in the two light curves from forming a significant minimum in the time-delay statistic, with Season~5 playing a particularly important role in anchoring all seasons close to the true delay. Excluding the 1.4-GHz results where there is little data and the intrinsic variability is particularly low gives an even narrower range of 40.5--46.9~d.

This range of delays broadly corresponds to the lower half of the \citetalias{koopmans00a} delay estimate, but falls completely outside (below) the 1~$\sigma$ range of the optical delay estimate. The results from the CSM and Pelt methods are very similar and usually agree to within a few days. A notable exception is the polarized flux density data for which the delays differ by 6.4~d.

\begin{figure}
\begin{center}
\includegraphics[width=\linewidth]{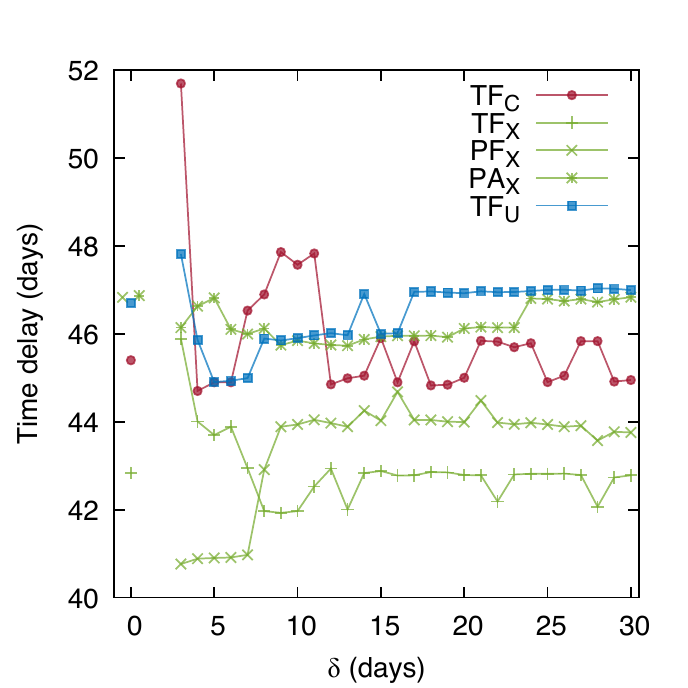}
\caption{Best-fit delay for the Pelt $D^2_{4,2}$ dispersion as a function of $\delta$ parameter for five of the 1600+434 datasets. The simpler $D^2_3$ statistic is shown along the $x = 0$ axis where the polarization results have been horizontally offset for clarity. For $\delta \la 12$~d the delays are relatively unstable, but higher values demonstrate a tendency for the delay to jump by $\pm$1~d. We do not show the 1.4~GHz results as these are not stable over any range of $\delta$. TF: total flux density, PF: polarized flux density.}
\label{fig:delta}
\end{center}
\end{figure}

Table~\ref{tab:delay} only shows the results for the simpler $D^2_3$ dispersion measure, but in Fig.~\ref{fig:delta} we also show the best-fit delays for the $D^2_{4,2}$ variant, for $3 \le \delta \le 30$~d. The $D^2_3$ results are also shown for comparison. The results demonstrate the following:
\begin{enumerate}
\item For low values of $\delta$ the delay is very unstable, presumably due to the number of included pairs being relatively small. The most common separation between epochs is 3~d and this is why we have not included $D^2_{4,2}$ results for $\delta < 3$~d.
\item For $\delta \ge 12$~d the delays become stable for all datasets but display a tendency to jump by $\pm$1~d. This is due to a regularity in the sampling of the data -- see below.
\item The stable values of delay are similar to those found using the much simpler $D^2_3$ statistic.
\item An exception is polarized flux density for which the delay is particularly unstable for $\delta < 9$~d. In the unstable region there is a tendency to congregate at delays close to the CSM result of 40.5~d.
\item The delay for the 1.4-GHz data (not shown) is a strong function of $\delta$ and given the paucity of the data, the low level of intrinsic variability and the prominent external variability in Season~4, we do not consider these data further for the purposes of determining the time delay.
\end{enumerate}
The $\pm 1$-d instability stems from the finite width ($\delta$) of the window used to include pairs. As described in Section~\ref{sec:obs}, the epochs tend to be separated by integer multiples of 1~d which means that varying the trial delay by 1~d tends to cause a significant change in the population of pairs that fall into the window. This in turn leads to regularly spaced inflection points in the dispersion spectrum, one of which is very likely to form the global minimum. The regular sampling also affects the nearest-neighbour variant, $D^2_3$, as the epochs that form neighbouring pairs also tend to change as the trial delay shifts by integer days.

A solution might be to use the $D^2_{4,3}$ dispersion variant which includes \textit{all} A, B pairs at every trial time delay and which therefore does not display this quantisation of delays. However, we find no range of the $\beta$ parameter (this statistic's equivalent of $\delta$) over which the best-fit delay is stable. As the two variants shown in Fig.~\ref{fig:delta} give consistent results for most of the datasets, we use the simplest $D^2_3$ for the remainder of the analysis.

A similar quantisation to that seen in the $D^2_{4,3}$ results also affects the CSM method. In this case, regularly spaced features in the $\chi^2$ spectrum occur due to bulk changes in points being interpolated as the trial delay is varied. For a point in one light curve, the pair of points in the other curve from which an interpolated value is calculated tends to change as the delay shifts by the spacing between the epochs i.e.\ by a multiple of 1~d. Taken over the ensemble of points, this leads to sudden changes in the $\chi^2$ spectrum spaced by 1~d. This is probably the reason why all of the CSM delays in Table~\ref{tab:delay} have a fractional part that lies between 0.3 and 0.7~d. Five out of six Pelt $D^2_3$ delays have fractional parts between 0.7 and 0.9~d.

We conclude that if the true delay happens to lie at or close to a maximum in the regularly spaced pattern that is imprinted on the CSM and Pelt spectra, it is unlikely to form the global minimum. More likely is that this forms at one of the directly neighbouring minima and thus we assume that there is $\pm$0.5~d systematic error that affects all delay determinations and which must be included in the final error budget.

\subsection{Radio light curve simulations}
\label{sec:sims}
  
\subsubsection{Total flux density}

In order to measure the delay and its associated uncertainty we followed a similar approach to that outlined by \citet{tewes13} for simulating light curves with external variability that is statistically equivalent to that seen in the real data.

\begin{figure*}
  \begin{center}
    \includegraphics[scale=0.95]{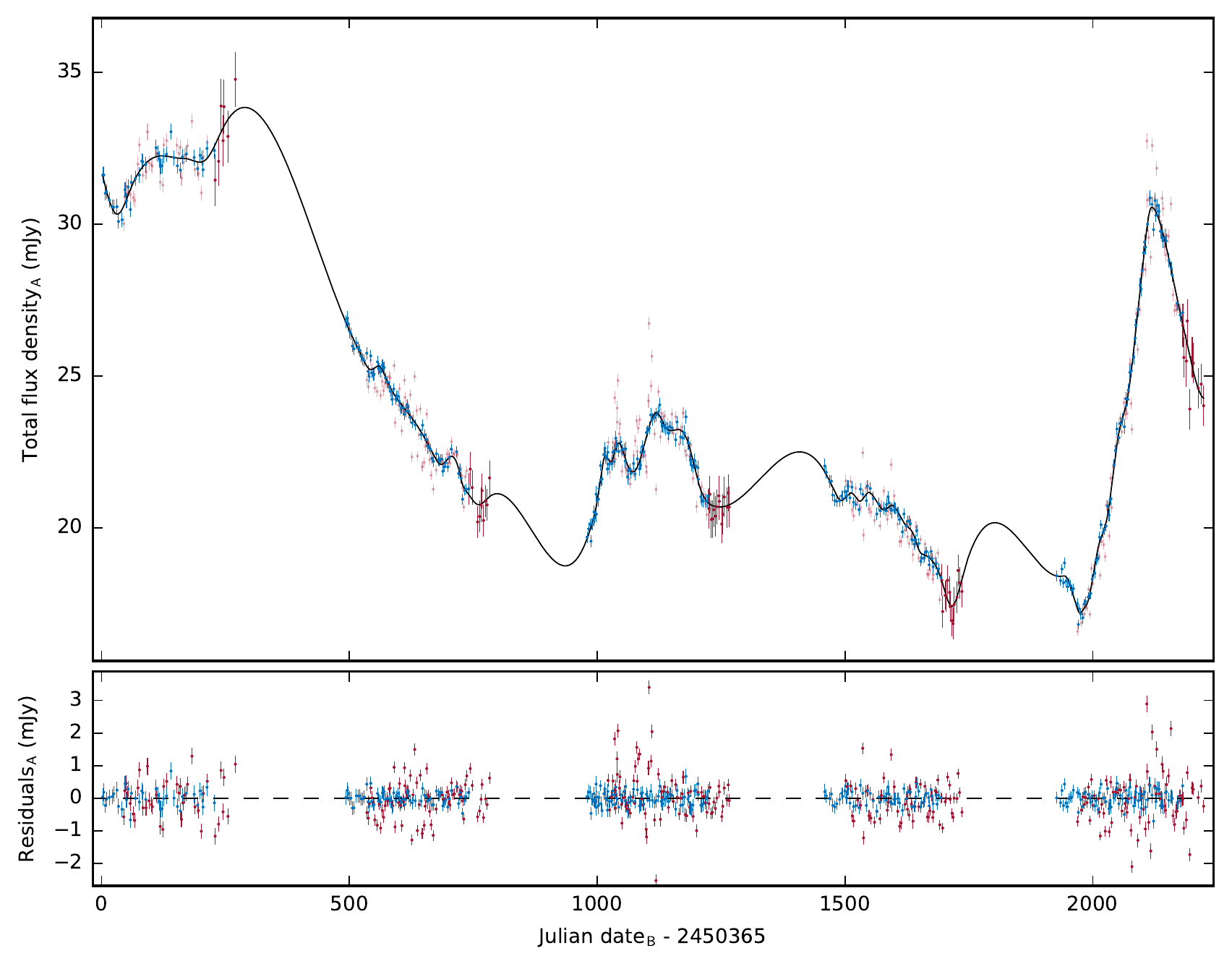}
    \includegraphics[scale=0.87]{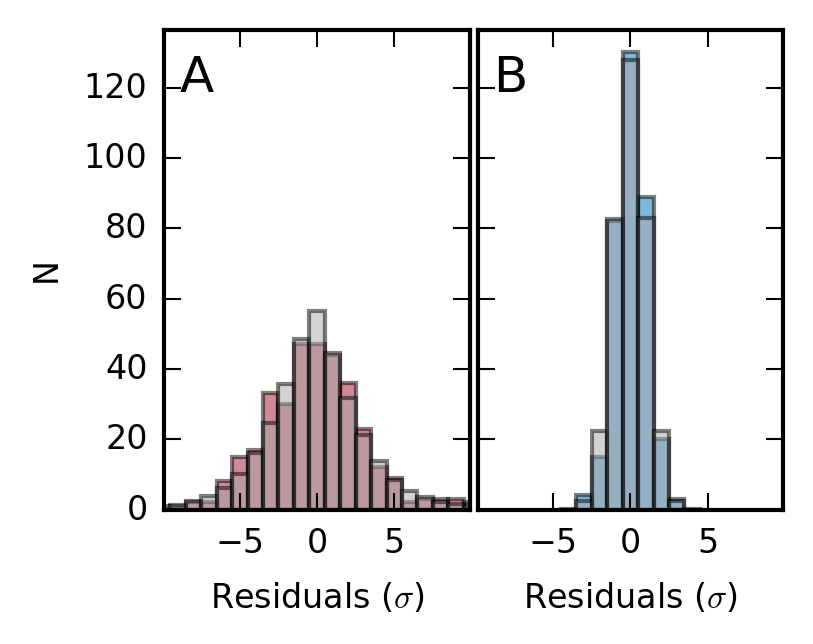}
    \includegraphics[scale=0.87]{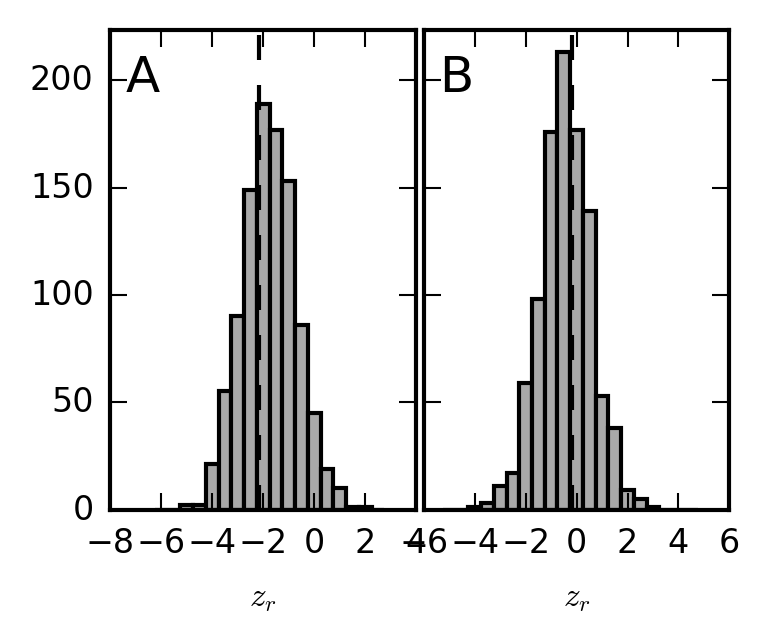}
    \includegraphics[scale=0.87]{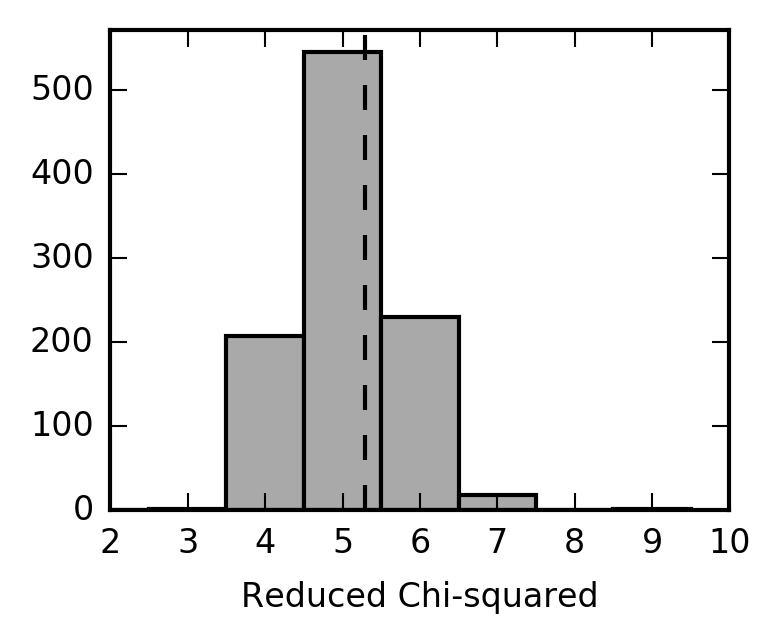}
    \caption{The top plot shows the 8.5-GHz total flux density data after the removal of the best-fit time delay and flux density ratio (first-order polynomial) found using the CSM method. The best-fitting spline was constrained using the image~B data and the image~A points at the end of each season -- these have larger error bars to avoid the external variability producing a spline with too many knots. The A data that was not used in the spline fitting is shown using fainter symbols. The middle panel shows the residuals around the spline fit -- the greater scatter of image~A is obvious. The bottom row illustrates that statistical properties of the simulated radio light curves are in good agreement with the real data. Left: histograms of the simulated residuals (grey) together with the real residuals (red and blue). Each residual has been divided by its associated error. Middle: histograms of the $z_r$ parameter. Right: histogram of $\chi^2_{\mathrm{dof}}$. In the last two cases the vertical dashed line shows the properties of the real data.}
    \label{fig:spline_x}
  \end{center}
\end{figure*}

The first step was to produce a model of the intrinsic quasar variability. This was done by fitting a cubic spline to the data after first removing the best-fit delay and $y$-offset. As the image~A data are strongly affected by external variability we exclude these from the fit over those time ranges where data from image~B is available. In the intra-season gaps the image~A data must be used but to make the spline fit here as smooth as possible we multiplied the error bars by the scatter in the image~A residuals and repeated the fit. In Fig.~\ref{fig:spline_x} this is demonstrated for the 8.4-GHz data -- the image~A data not used in the fit are plotted more faintly, whilst those used as constraints are shown with the larger error bars. The spline contains 68 knots with an average separation of 17~d. The greater magnitude of the image~A residuals is obvious.

One of the statistical tests used by \citet{tewes13} is the number of `runs' in the residuals i.e.\ the number of sequences where consecutive values are either positive or negative, usually parameterised as a deviation in sigmas from what would be expected for randomly distributed data ($z_r$). Whilst this should be equal to or less than zero \citep{wall03}, for image~B we found $z_r \ga +$5~$\sigma$ at 8.4 and 15~GHz. This perhaps indicates that the errors have been underestimated at these frequencies as increasing them by 40~per~cent produced a smoother spline, residuals that were Gaussian-distributed with a width of 1~$\sigma$ and which had the expected number of runs. That we might have underestimated the errors at the higher frequencies is quite plausible given that flux calibration becomes more difficult as the frequency increases. As we prefer to overestimate rather than underestimate the uncertainties, we have used the larger error bars throughout this paper.

The next stage is to derive simulated residuals for A and B. For image~A, the basis of these is a random time series generated using the power-law-noise algorithm of \citet{timmer95}, the characteristics of which are controlled by four parameters: the amplitude ($A$), the slope of the frequency spectrum ($\beta$) and the highest and lowest frequencies present ($f_{\mathrm{max}}$ and $f_{\mathrm{min}}$). For all frequencies we set $f_{\mathrm{max}} = 5$ and $f_{\mathrm{min}} = 1 / 200$~d$^{-1}$.

To reflect the changing characteristics of the external variability with time, at 5 and 8.5~GHz we measured the rms of the residuals of image~A using a sliding window with a width of seven epochs and used this to scale the amplitude of the time series. This is particularly important in order to simulate an event similar to the `caustic crossing' in Season~3. At 15~GHz, this rms scaling was instead done using a season average. As a final modification, for all frequencies we also measured the mean of the residuals per season and offset each season's time series such that the simulated residuals had the same mean.

The values of $A$ and $\beta$ were then set by trial and error until they produced simulated light curves with the same statistical properties (magnitude of residuals around a spline fit and number of runs) as the real data. For image~B, for each epoch we selected a random residual from a Gaussian distribution with a $\sigma$ equal to that epoch's uncertainty. In all cases 1000 simulated radio light curves were generated.

To create the actual simulated radio light curves the residuals were applied to the spline that represents our model of the intrinsic variability. In doing so, we constructed light curves with time delays different to that used to form the combined curve by sampling the spline at $t_i + \tau + \delta \tau_{\mathrm{A}}$ and $t_i + \delta \tau_{\mathrm{B}}$ where $t_i$ represents the original timestamps of each epoch $i$, $\tau$ is the time delay used to form the combined variability curve and $\delta \tau_{\mathrm{A}}$ and $\delta \tau_{\mathrm{B}}$ are random offsets centred on 0~d with a maximum value of e.g.\ $| \delta \tau_{\mathrm{A}}| = 3$~d.

The bottom panel of Fig.~\ref{fig:spline_x} shows the results of this process applied to the 8.5-GHz total flux density data with the CSM technique. Histograms of the residuals themselves (the residuals from the real data and the average of the 1000 simulated residuals), the number of runs $z_r$ and $\chi^2_{\mathrm{dof}}$ found in each simulation are shown together with the equivalent values measured from the real data. In all cases a good match has been possible.

\subsubsection{Polarization}

The relatively low $\chi^2_{\mathrm{dof}}$ values found for the polarized flux density and EVPA data suggest that these do not contain significant levels of external variability. We therefore initially performed a spline fit using all the A and B data, but smoothing the residuals (which are much noisier than for the total flux density data - see Section~\ref{sec:external}) revealed that external variability is also affecting the polarization properties of image~A.

We therefore followed the same procedure as with total flux density and constrained the spline fit using the image~A data in the gaps between seasons and the image~B data otherwise. A satisfactory spline fit required a 10~per~cent reduction in the magnitude of the polarized flux density and EVPA uncertainties and again this has been applied throughout this paper. One difference was the use of a quartic spline due to the greater magnitude of the polarization variability. The number of knots in the polarized splines was similar to that found for the 8.5-GHz total flux density data, 64 and 71 (cf. 68) for polarized flux density and EVPA respectively.

\subsection{The final time delays and their uncertainties}
\label{sec:finaldelay}

For each dataset we measured the difference between the input and output delay of the 1000 simulations, $\Delta \tau$. From this we then measured the median ($\Delta \tau_{\mathrm{med}}$) and the upper and lower boundaries of the central 68.3~per~cent of the distribution. The difference between the median and the two boundaries gives the plus and minus 1-$\sigma$ random statistical error, $\sigma_{\mathrm{random}}$. These values are shown in Table~\ref{tab:tcomp}.

\begin{table*}
  \centering
  \caption{Results from the simulated radio light curves. For both the CSM and Pelt methods we report the median difference between the input and output delays ($\Delta \tau_{\mathrm{med}}$), the 1-$\sigma$ width of the delay distribution ($\sigma_{\mathrm{random}}$), the total systematic error ($\sigma_{\mathrm{sys}}$) and the delay ($\tau$) together with its total uncertainty ($\sigma_{\mathrm{tot}}$). The delays and their total uncertainties are plotted in Fig.~\ref{fig:tcomp}. TF: total flux density, PF: polarized flux density.}
  \label{tab:tcomp}
  \begin{tabular}{cc@{\hskip 1.0cm}cccc@{\hskip 1.0cm}cccc} \\ \hline
    Data & $\nu$ (GHz) & \multicolumn{4}{c}{CSM Delay (d)} & \multicolumn{4}{c}{Pelt Delay (d)} \\
    &  & $\Delta \tau_{\mathrm{med}}$ & $\sigma_{\mathrm{random}}$ & $\sigma_{\mathrm{sys}}$ & $\tau \pm \sigma_{\mathrm{tot}}$ & $\Delta \tau_{\mathrm{med}}$ & $\sigma_{\mathrm{random}}$ & $\sigma_{\mathrm{sys}}$ & $\tau \pm \sigma_{\mathrm{tot}}$ \\ \hline
    TF   & 5   & $-$2.3 & $^{+4.2}_{-3.5}$ & $\pm{2.4}$ & $44.6\pm{4.8}$ & 0.1    & $^{+3.4}_{-5.0}$ & $\pm{0.5}$ & $45.3\pm{5.0}$ \\
    TF   & 8.5 & 0.0    & $^{+2.0}_{-1.8}$ & $\pm{0.5}$ & $42.3\pm{2.1}$ & 0.6    & $^{+2.3}_{-2.0}$ & $\pm{0.8}$ & $42.8\pm{2.4}$ \\
    PF   & 8.5 & $-$0.2 & $^{+3.6}_{-3.2}$ & $\pm{0.5}$ & $40.5\pm{3.6}$ & 0.7    & $^{+3.5}_{-3.1}$ & $\pm{0.9}$ & $46.9\pm{3.6}$ \\
    EVPA & 8.5 & 0.1    & $^{+2.4}_{-2.2}$ & $\pm{0.5}$ & $45.5\pm{2.5}$ & 0.0    & $^{+2.5}_{-2.6}$ & $\pm{0.5}$ & $44.9\pm{2.6}$ \\
    TF   & 15  & $-$0.1 & $^{+2.1}_{-2.2}$ & $\pm{0.5}$ & $44.5\pm{2.3}$ & $-$0.8 & $^{+2.6}_{-2.3}$ & $\pm{0.9}$ & $46.7\pm{2.8}$ \\ \hline
  \end{tabular}
\end{table*}

For six of the ten data/minimisation-technique combinations shown in Table~\ref{tab:tcomp}, $|\Delta \tau_{\mathrm{med}}| \le 0.2$~d and thus very small compared to $\sigma_{\mathrm{random}}$. Most of the remaining values fall in the range $0.6 \le |\Delta \tau_{\mathrm{med}}| \le 0.8$, whilst the 5-GHz CSM value of $\Delta \tau_{\mathrm{med}} = -2.3$~d is much larger.

The origin of these biases is not entirely clear, but could be a consequence of the external variability. It is frequently observed that the trial time delay correlates with the $y$-offset (this is most easily understood for a purely linear variation in e.g.\ total flux density) and if the flux ratio is distorted by the presence of external variability, this could bias the time delay. This highlights the importance of both realistically simulating the external variability when performing the error analysis and creating variability curves with different input time delays.

If, however, the non-zero values of $\tau_{\mathrm{med}}$ reflected a bias in the value of the time delay found by the CSM or Pelt analysis, it would be expected that correcting for the bias would produce better agreement between the various datasets. This is not observed. For example, subtracting the bias from the 5-GHz results (both CSM and Pelt, although for the latter the bias is only 0.1~d) makes them less consistent, the corrected values being equal to 48.7 and 45.5~d. Of the four datasets where a large ($>$0.2~d) median offset is seen, the delays are in better agreement in two cases.

Regardless of the origin of the bias, it is clearly prudent to include it in the error budget and we have interpreted $\Delta \tau_{\mathrm{med}}$ as a systematic error, $\sigma_{\mathrm{sys,1}}$. Another systematic effect, $\sigma_{\mathrm{sys,2}}$, is the $\pm$0.5~d time-delay quantisation described in Section~\ref{sec:mseasonres} that is caused by the semi-regular sampling of the observing epochs. As the simulations used the same sampling as the real data, the quantisation effect should not be responsible for the non-zero median delays and thus we treat both systematic effects as independent and form a total systematic error, $\sigma_{\mathrm{sys}}$, by adding them in quadrature.

Other sources of systematic error should be considered. Is there, for example, a systematic offset between the Pelt and the CSM methods? The results for the individual seasons (Table~\ref{tab:delay}) can be used to investigate this as there are 50 separate time-delay measurements, 25 from each method. The difference between the delays found for each datset, $\Delta\tau_{\mathrm{Pelt-CSM}}$, is positive in 12 cases and negative in a further 12 -- in one case the delay is the same. This highly symmetric distribution strongly suggests that there is no systematic error introduced by the choice of delay-search technique and we have therefore not included any such contribution.

\begin{figure}
\begin{center}
\includegraphics[width=\linewidth]{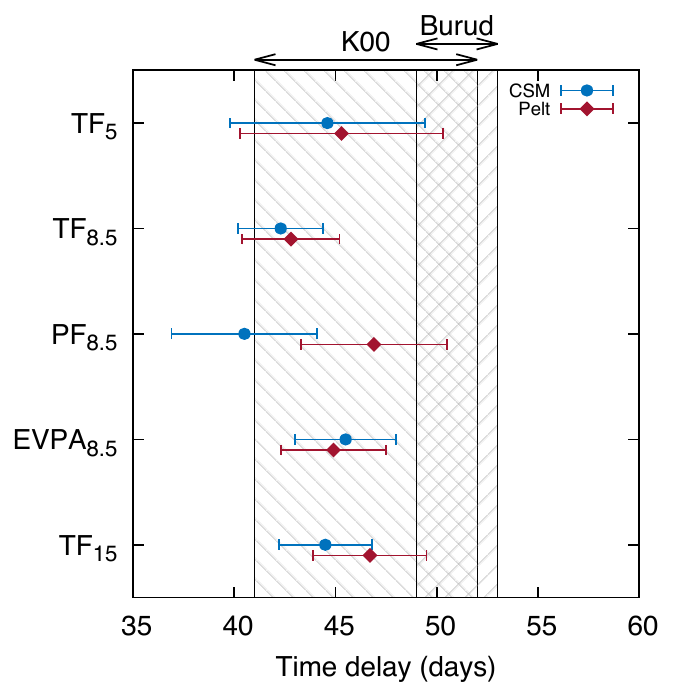}
\caption{Time delays and associated 1-$\sigma$ uncertainties for all five datasets. The hatched regions show the 1-$\sigma$ error range of the previously published radio \citepalias{koopmans00a} and optical \citep{burud00} analyses. The new delay values presented here are all consistent with one another but are generally lower than the optical time delay. TF: total flux density, PF: polarized flux density.}
\label{fig:tcomp}
\end{center}
\end{figure}

The final step is to combine the random and systematic errors into a total error estimate for each delay. In general, none of the random delay distributions are perfectly Gaussian, but they are nearly symmetric in most cases. We have therefore formed a final error estimate, $\sigma_{\mathrm{tot}}$, by taking the larger of the upper and lower values of $\sigma_{\mathrm{random}}$ and combining this in quadrature with $\sigma_{\mathrm{sys}}$. The final time delays and their uncertainties are shown in Table~\ref{tab:tcomp} and plotted in Fig.~\ref{fig:tcomp}.

\section{Discussion}
\label{sec:discussion}

\subsection{The time delay}

We have presented time delays from five independent datasets using two different techniques. The delays from each dataset are consistent with one another at the 1-$\sigma$ confidence level, as are the results from the CSM and Pelt techniques. This indicates that all significant sources of random and systematic error have been accounted for.

In principle, a more accurate delay estimate would be possible by combining the delays from the different datasets, but this is complicated by the sometimes non-Gaussian nature of the errors plus the likelihood that the systematic errors, especially those relating to the semi-regular spacing of the observing epochs, are correlated between different datasets. A further complication is that the different datasets might not be measuring the same delay due to e.g.\ frequency-dependent offsets caused by a shift in the position of the core of the radio jet due to synchroton self-absorption \citep[e.g.][]{kovalev08}. The centroids of the polarization and total flux density brightness distributions might also differ. Spatial offsets in the source plane will in general correspond to different time delays \citep[e.g.][]{cheung14} and will be a further source of systematic error that is difficult to estimate.

The nominally most accurate delay is $42.3 \pm 2.1$~d from the CSM analysis of the 8.5-GHz total flux density and we note that in general the systematic errors are smallest for the CSM results. Combining the three CSM delays with the smallest error results in a combined delay estimate of $43.9 \pm 1.3$~d. Combining all five CSM results would give $43.6 \pm 1.2$~d. Ultimately, although the combined results give slightly higher delays, both are consistent with the total flux density 8.5-GHz value. In order to be as statistically rigorous as possible, we therefore take the 8.5-GHz total flux density value as our best estimate of the delay.

In comparison with the previous time-delay estimations in this lens system, our results are all lower than the optical value. The only radio delays that are compatible with the optical result at 1~$\sigma$ are those with the largest uncertainties and thus we conclude that the optical delay is too high. As with the potential frequency-dependent radio delays already mentioned in Section~\ref{sec:finaldelay}, different values of radio and optical delays could result from offsets between the position of the optical and radio emitters. However, \citet{kovalev08} determine an average shift between the optical and 8.6-GHz positions of quasars that is an order of magnitude smaller than that between 8.6~GHz and lower radio frequencies. We therefore consider it more likely that the microlensing present in the optical data has led to the delay being overestimated and/or the error underestimated.

In contrast, our new delays are compatible with the previously published radio delay of \citetalias{koopmans00a}. As with the optical delay, the previous radio delay seems to have been overestimated, but its 1-$\sigma$ uncertainty encompasses the new results. The reason for the previous overestimate is likely to have been the assumption of a flux density ratio that was too high. \citetalias{koopmans00a} used a value of 1.212 whereas we find values of 1.191 and 1.194 from the CSM and Pelt analyses of these data (Season~1). However, as it is now known that the flux density ratio slowly varies as a function of time, the entire exercise of using a flux ratio from one season to constrain the time delay in another is inevitably prone to error.

Taking the 8.5-GHz result as our most reliable delay, the radio time delay has reduced in this system by 10~per~cent.

\subsection{The nature of the external variability}
\label{sec:external}

\begin{figure*}
\begin{center}
\includegraphics[width=\linewidth]{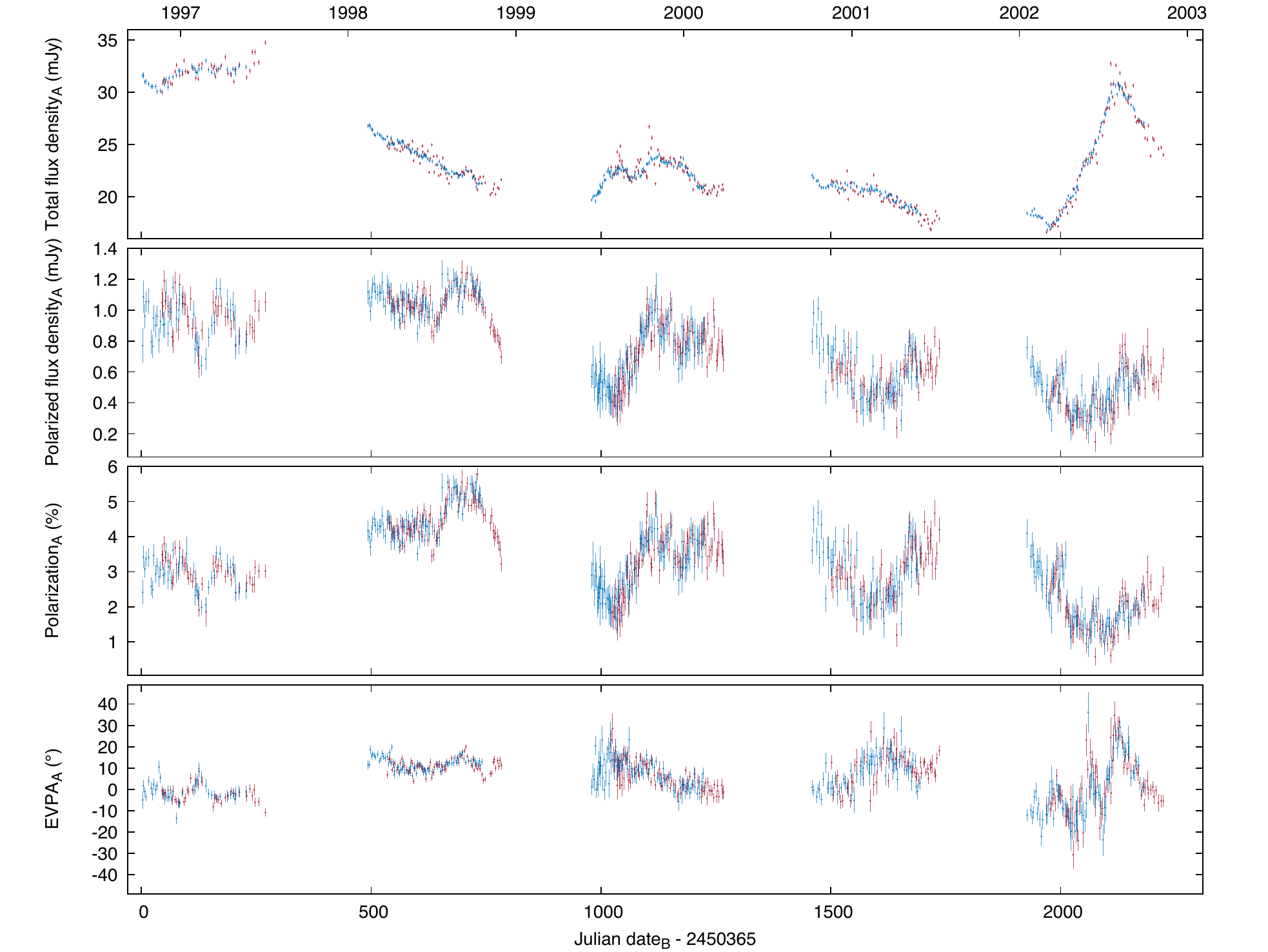}
\caption{VLA 8.5-GHz variability curves of B1600+434 after removal of the time delay (42.3~d) and the $y$-offsets. From top to bottom: total flux density, polarized flux density, percentage polarization and polarization position angle. As image A has been shifted along the $x$-axis, all time labels refer to the observation of image B. Similarly, $y$-offsets have been applied to image~B so as to agree with the image~A values.}
\label{fig:x_shift}
\end{center}
\end{figure*}

\begin{figure*}
\begin{center}
\includegraphics[width=\linewidth]{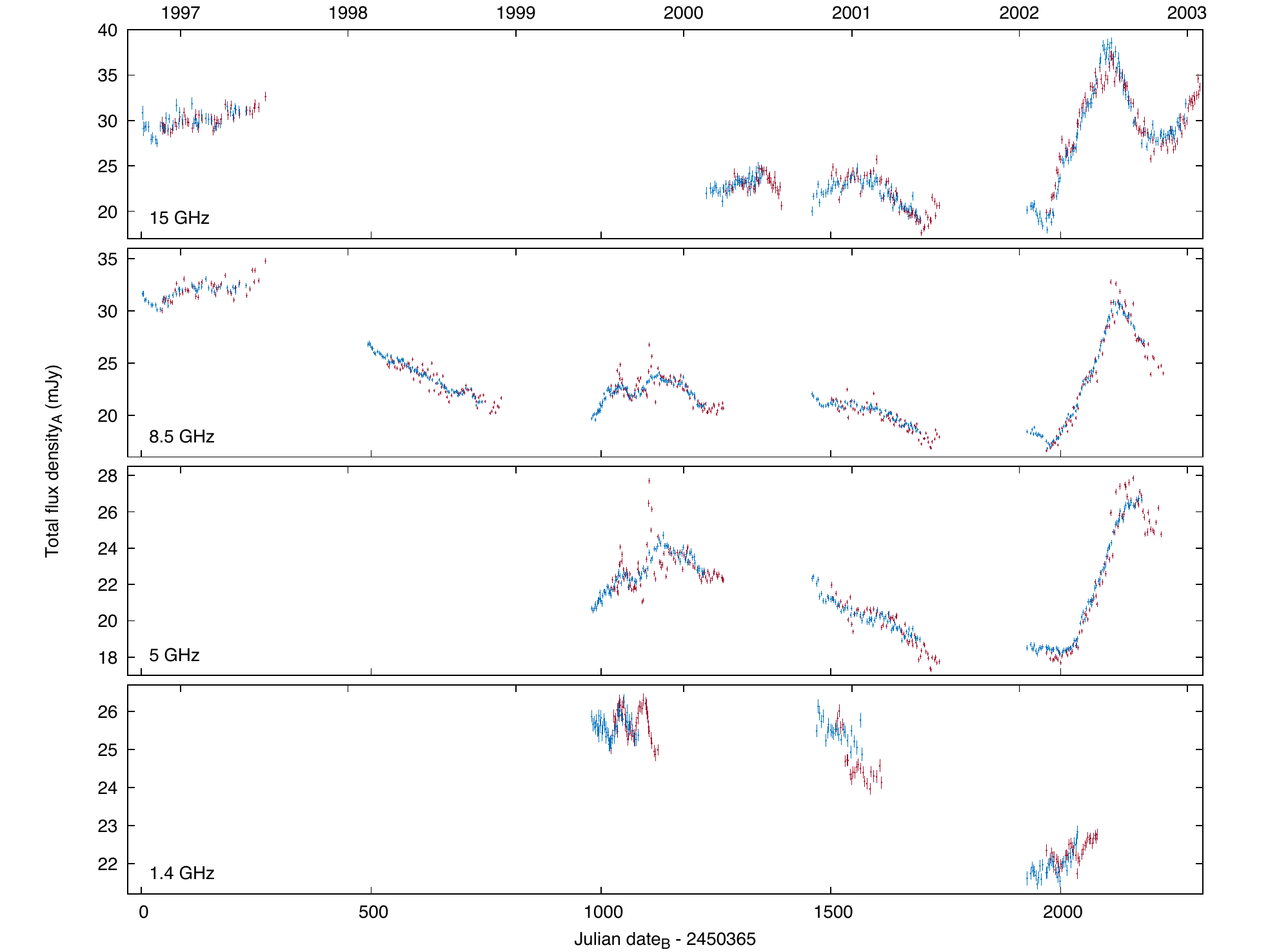}
\caption{VLA total flux density variability curves of B1600+434 after removal of the time delay (42.3~d) and the $y$-offsets. As image A has been shifted along the $x$-axis, all time labels refer to the observation of image B. Similarly, $y$-offsets have been applied to image~B so as to agree with the image~A values.}
\label{fig:tf_shift}
\end{center}
\end{figure*}

\begin{figure*}
\begin{center}
\includegraphics[width=\linewidth]{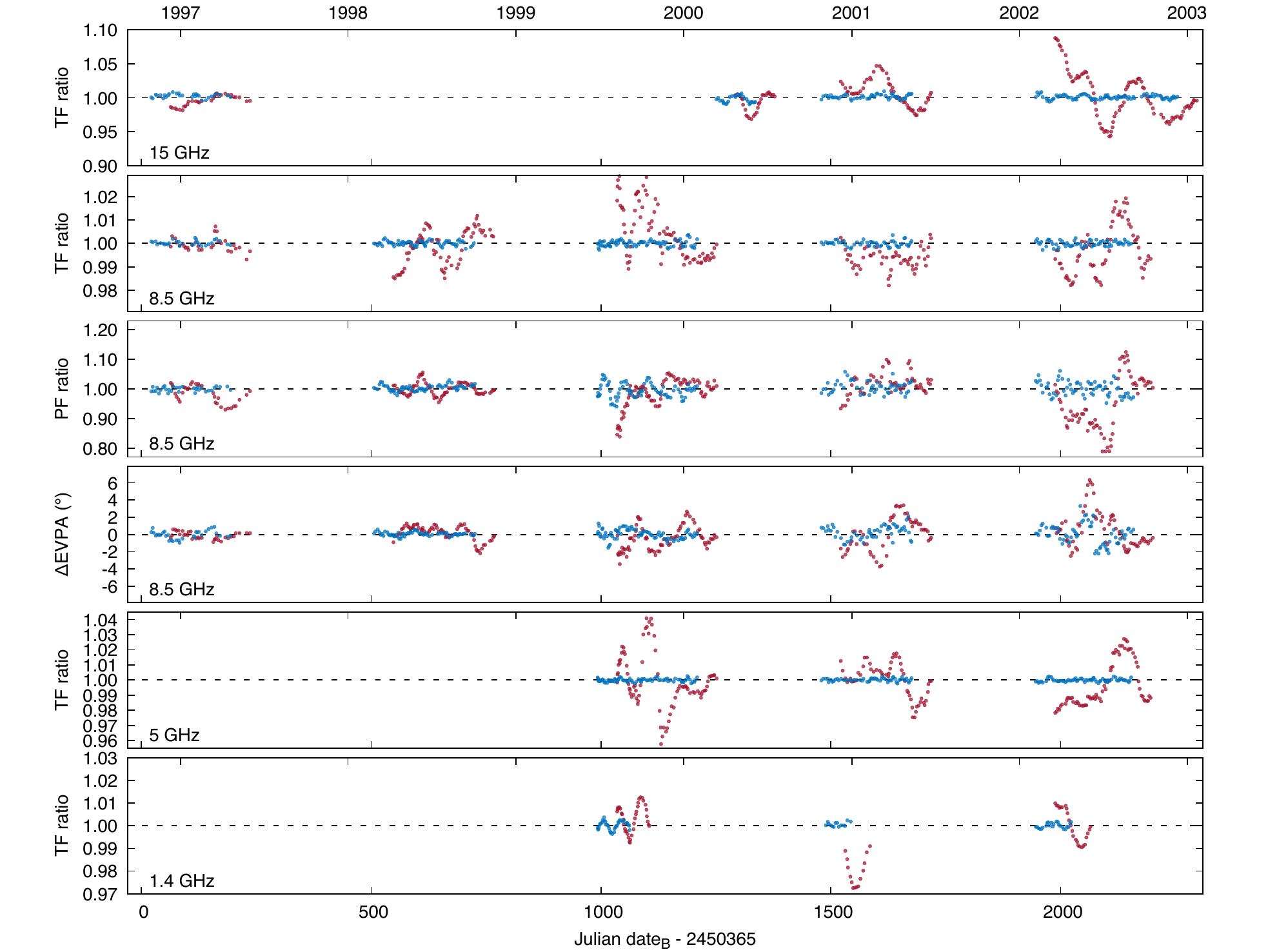}
\caption{Total (TF) and polarized (PF) flux density ratios and EVPA residuals for A (red) and B (blue). The data have been smoothed using a box filter with a width of 11 epochs and the ten epochs at the beginning and end of each season are not shown. Note the magnitude of the polarized flux variations which are much larger than those seen in total flux density at the same frequency (8.4~GHz). The long-term linear variation of the A/B total flux density ratio has been removed.}
\label{fig:residuals}
\end{center}
\end{figure*}

Plots of all data with the time delay and $y$-offsets removed are shown in Figs~\ref{fig:x_shift} (8.5~GHz total flux density and polarization) and \ref{fig:tf_shift} (all total flux density data). In all cases the 8.5-GHz total flux density delay has been used (42.3~d) and the $y$-offsets optimized for this. The main benefit of plotting the data in this way is to highlight the external variability in image~A. This is made even clearer in Fig.~\ref{fig:residuals} where we show the shifted and scaled A and B data with the intrinsic variability removed (the A and B data are divided by the spline fit for total and polarized flux density and the spline is subtracted for EVPA) for all data. In order to increase the SNR, especially for the polarization data, we have smoothed the data using a box filter with a width of 11~epochs ($\sim$one month) and excluded the ten epochs at the beginning and end of each season.

\citet{koopmans00b} concluded that the most likely explanation for the external variability is microlensing of superluminal jet components by compact objects in the halo of the lensing galaxy. Key to distinguishing between this possibility and the competing hypothesis of scintillation is the fractional rms variability (modulation index $m$) which for microlensing should decline with decreasing frequency. Although lensing is inherently achromatic, the larger source sizes expected at lower frequencies will lead to the variations caused by multiple caustics to be averaged. Simultaneous WSRT monitoring at 1.4 and 5~GHz demonstrated that $m_{\mathrm{A+B}}$ (the two images are unresolved at these frequencies) is indeed smaller at the lower frequency, although there additional reasons for disfavouring scintillation, notably the time-scales of the variability. External variability in image~B was not ruled out and assumed to be present but at a lower level than in A.

In Table~\ref{tab:modidx} we show the value of $m$ for each image during Seasons~4 and 5. A measurement using all the data at each frequency would not give a fair comparison due to the different time periods covered and these two seasons have been selected as they are the only ones with data at all four frequencies. The fractional rms variability has been calculated from the A and B data after dividing by the spline fit and without any smoothing, but we do not use any A data where this was used to constrain the spline fit (e.g.\ in the gaps between seasons) as this would lead to an underestimate of the external variability in this image.

With only two images to compare, determining the true level and nature of external variability in image~B is difficult, but we find no evidence that we are seeing anything other than the intrinsic variation of the lensed quasar in this image which varies remarkably smoothly throughout the monitoring campaign. Short-term variability, on a time-scale of several epochs, appears to be completely absent and there is no evidence at all of any events similar to e.g.\ the two sudden increases in the flux density of image~A during Season~3.  The modulation index of this image appears to hardly vary with frequency and essentially reflects the measurement errors, although this is somewhat expected given that the spline was constrained using the image~B data only. Fitting instead a linear function to the 8.5-GHz data from Season~2 allows a direct comparison with the \citet{koopmans00b} modulation index for these data and we find a smaller value of $m_{\mathrm{B}} = 1.1$ cf. 1.6~per~cent.

As there is no reason to believe that image~B is affected by external variability, we therefore assume that its observed modulation index is purely due to random measurement error. This can then be used to form a corrected value of $m_{\mathrm{A}}$ where
\begin{equation}
m_{\mathrm{A,cor}} = \sqrt{m_{\mathrm{A}}^2 - m_{\mathrm{B}}^2}.
\end{equation}

\begin{figure}
\begin{center}
\includegraphics[width=\linewidth]{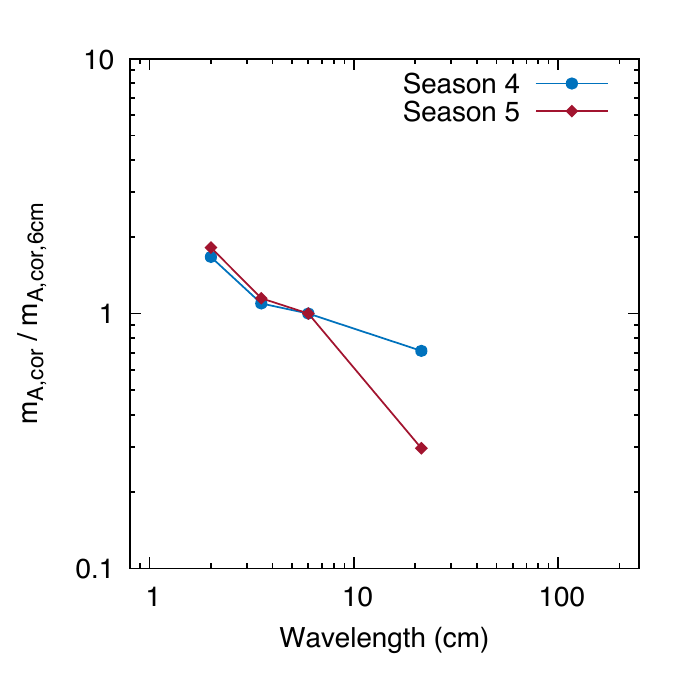}
\caption{Modulation index for image~A as a function of wavelength -- see Table~\ref{tab:modidx}. The data from each season are normalised to the 6-cm value. In both seasons there is a continuous decline in $m$ with wavelength which is the expected behaviour for microlensing. This figure should be compared to fig.~11 of \citet{koopmans00b}.}
\label{fig:modidx}
\end{center}
\end{figure}

In Fig.~\ref{fig:modidx} we show $m_{\mathrm{A,cor}}$ as a function of wavelength and normalised to the 6-cm value i.e.\ following fig.~11 of \citet{koopmans00b}. This demonstrates that for both Season~4 and 5 the modulation index declines uniformly with increasing wavelength and thus we conclude that the new multi-frequency data are entirely consistent with microlensing being the origin of the external variability in image~A. The modulation index appears to be less stable at the lowest frequency of 1.4~GHz, but we caution that there is less data here due to the poor angular resolution in the smaller VLA configurations. Season~3 also demonstrates very low external variability at 1.4~GHz with $m_{\mathrm{A,cor}} = 0.7$~per~cent. 

\begin{table*}
  \centering
  \caption{Modulation index for image~A as a function of wavelength. Results are shown for Seasons~4 and 5 as these contain data at all four frequencies and do not include the third season's `caustic-crossing event' which is only fully sampled in both images at 5 and 8.5~GHz. The `A,cor' values have had the contribution from B removed. These data are plotted in Fig.~\ref{fig:modidx}.}
  \label{tab:modidx}
  \begin{tabular}{ccccccc} \\ \hline
    & \multicolumn{3}{c}{Season 4} & \multicolumn{3}{c}{Season 5} \\
    Frequency (GHz) & $m_\mathrm{A}$ (\%) & $m_\mathrm{B}$ (\%) & $m_\mathrm{A,cor}$ (\%) & $m_\mathrm{A}$ (\%) & $m_\mathrm{B}$ (\%) & $m_\mathrm{A,cor}$ (\%) \\ \hline
    1.4 & 1.7 & 0.8 & 1.5 & 1.1 & 0.8 & 0.8 \\
    5   & 2.2 & 0.8 & 2.1 & 2.8 & 0.7 & 2.7 \\
    8.5 & 2.4 & 0.7 & 2.3 & 3.1 & 0.6 & 3.1 \\
    15  & 3.8 & 1.5 & 3.5 & 5.1 & 1.4 & 4.9 \\ \hline
  \end{tabular}
\end{table*}

The smoothed data in Fig.~\ref{fig:residuals} demonstrate that at 8.4~GHz the external variability of the polarized flux density is considerably greater than that of the total flux density, with a maximum magnitude of $\sim$20 cf. 3~per~cent. Differences between polarized and total flux density are expected given that the polarization structure of jets tends to be different to that seen in total flux density, for example that the core polarization of quasars tends to be rather low ($<$5~per~cent) and increase along the jet \citep{lister05}. Within the context of microlensing, the greater variability suggests less spatial averaging over a caustic structure, perhaps due to the polarization covering a smaller total area. The EVPA variations are small which suggests that this changes very little along the jet.

\section{Improving the lens model}
\label{sec:lensmodel}

Although the time delay in this system has been refined, its use for $H_0$ determination is currently limited due to a combination of the small number of constraints available from only two images together with the complicated lensing mass distribution which includes a galaxy group and one particularly prominent nearby perturber. Ideally, additional constraints on the lens model could be found and the large amount of data taken as part of these monitoring campaigns can potentially also help on this front.

\subsection{Stacking the VLA data}

As the flux densities of both lensed images have been measured at each epoch, it is possible to subtract them from the $u,v$ data and then combine the epochs to make a very sensitive continuum image with which a search can be made for any residual lensed emission. The subtraction was done in \textsc{aips} using \textsc{uvmod} which also allows the subtraction of the Stokes $Q$ and $U$ flux densities. Once every epoch of a particular monitoring season had been subtracted, these were combined into a single dataset using the \textsc{stuffr} procedure. This converts the timestamps of the data to hour angle and then averages the data from each epoch together per baseline -- this leads to a much smaller dataset than would otherwise be the case. The averaged data from each season were then combined using \textsc{dbcon}.

\begin{figure*}
\begin{center}
\includegraphics[scale=0.4]{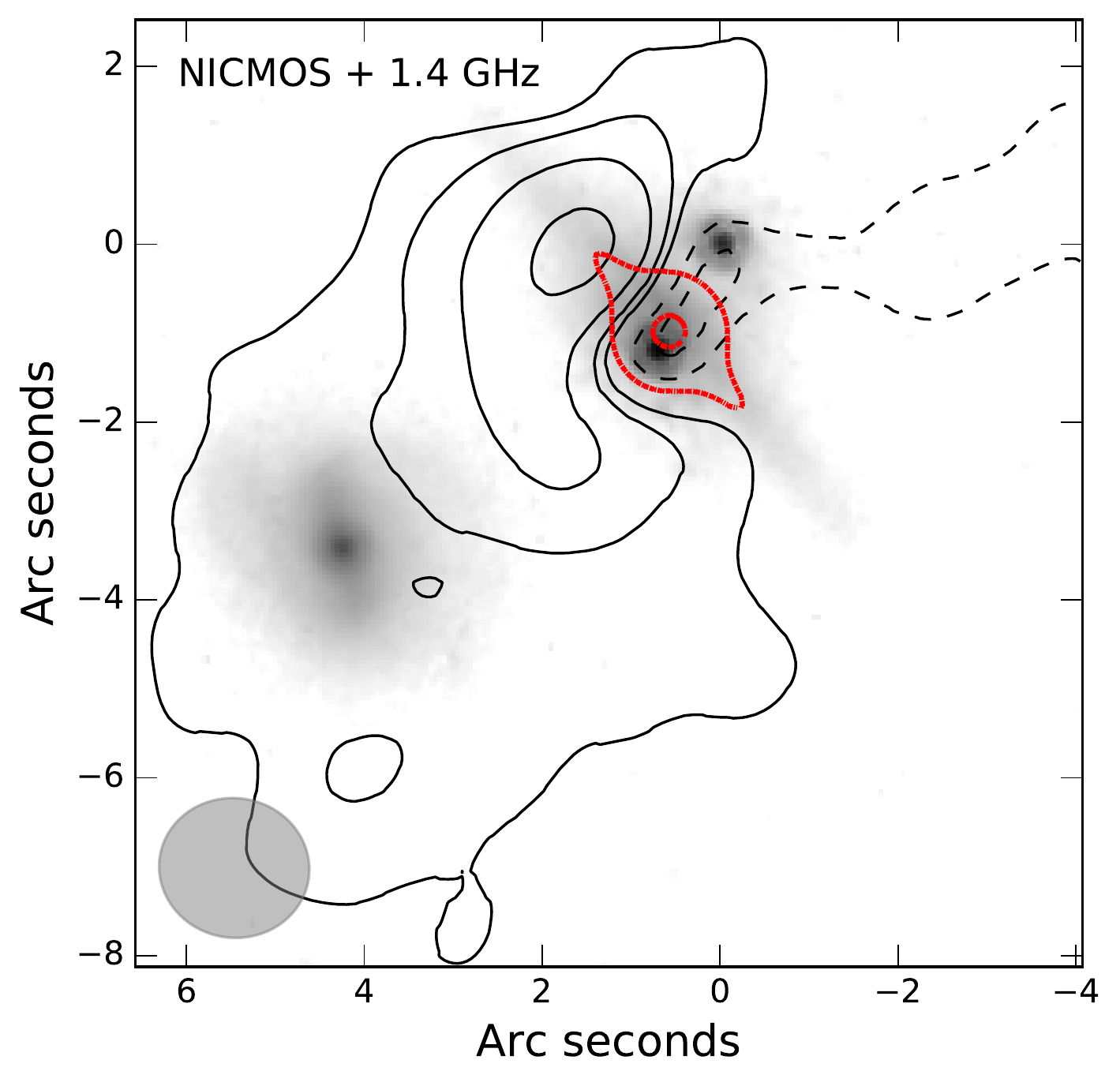}
\includegraphics[scale=0.4]{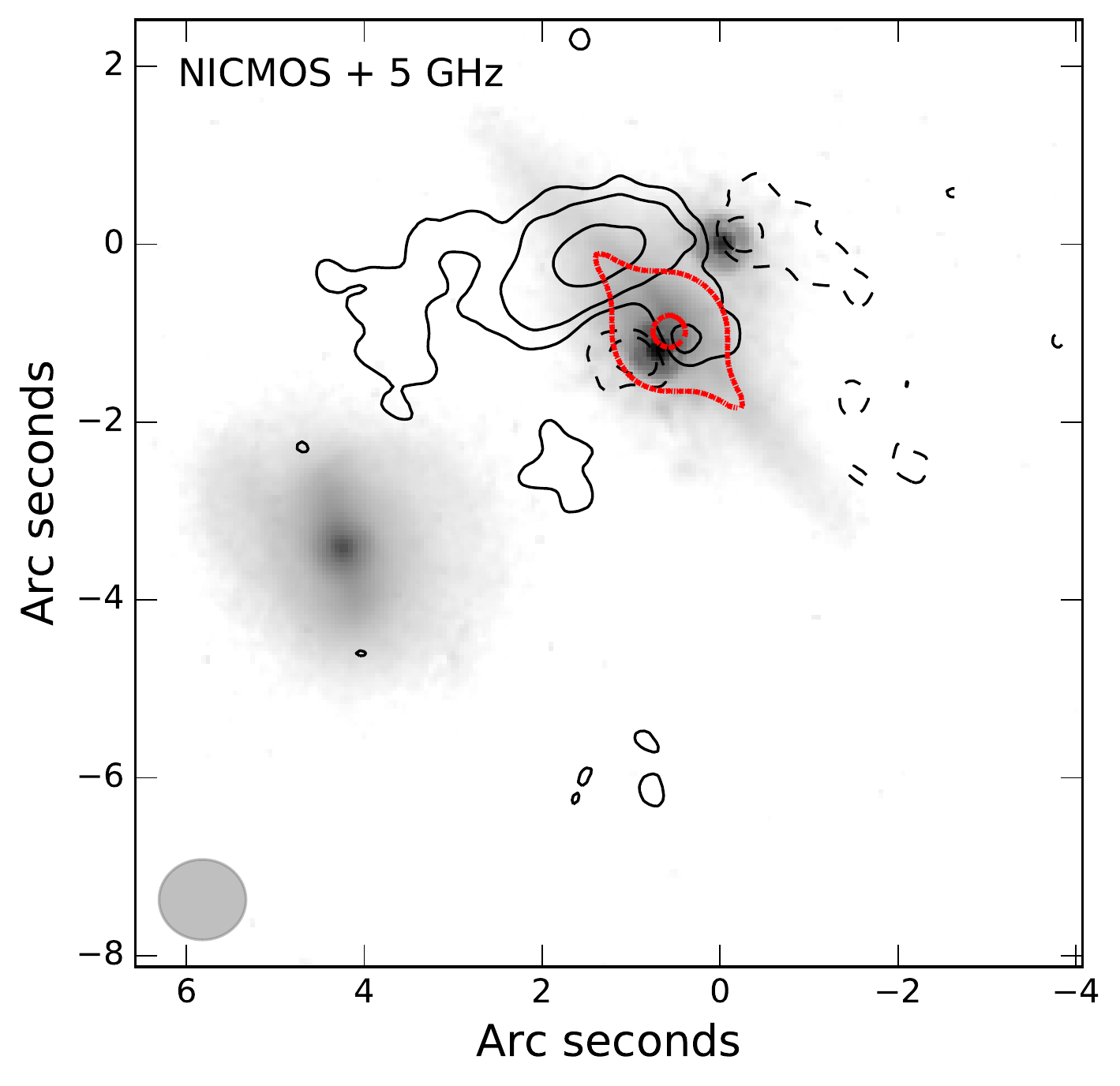}
\includegraphics[scale=0.4]{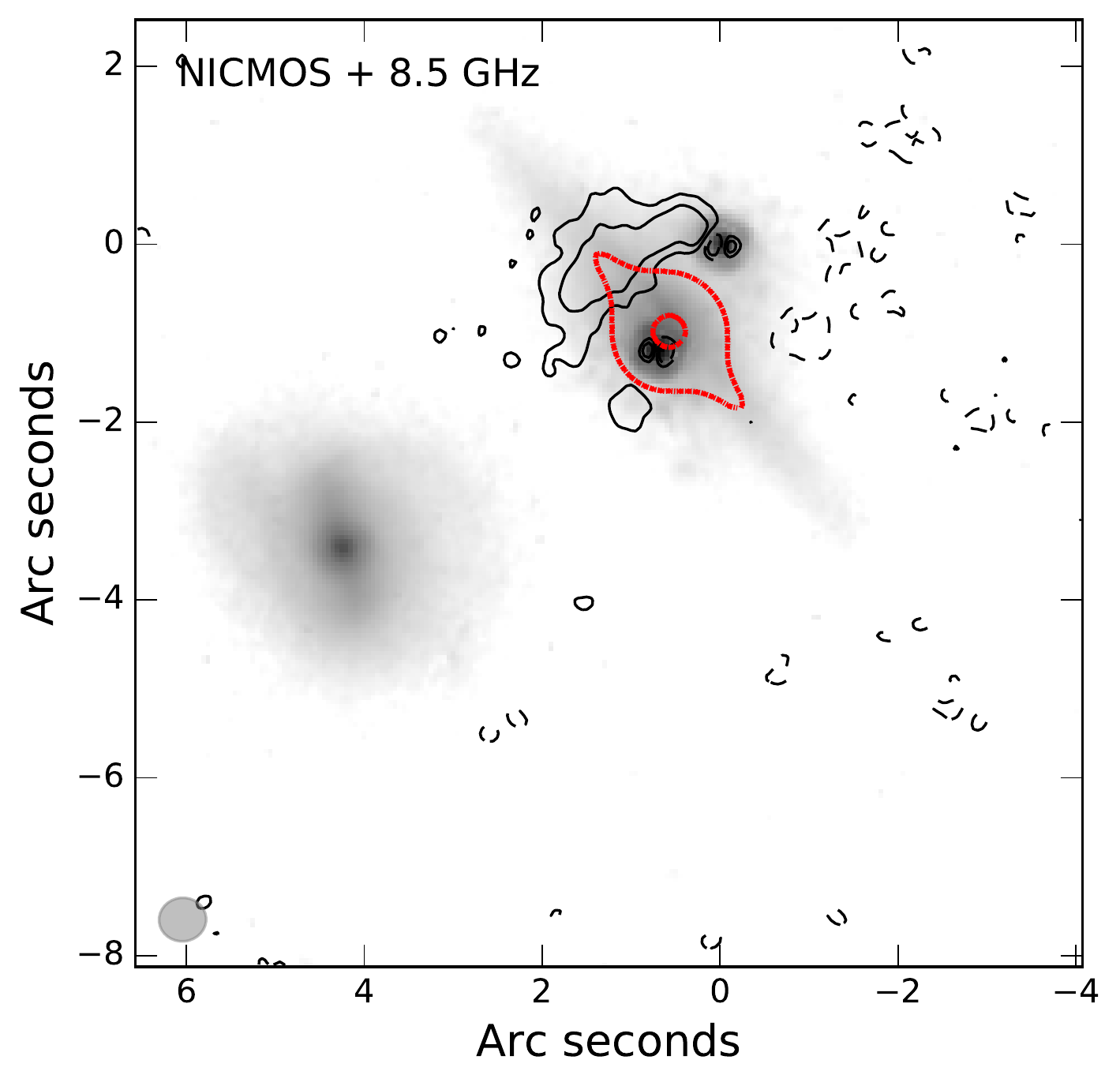}
\caption{Contours: Combined VLA maps of 1600+434 after subtraction of the emission from the lensed images A and B. From left to right: 1.4~GHz, 5~GHz and 8.5~GHz. Contours are plotted at $-2$, $-1$, 1, 2, 4 times the 3-$\sigma$ rms noise where $\sigma_{\mathrm{1.4~GHz}} = 9.2$, $\sigma_{\mathrm{5~GHz}} = 6.0$ and $\sigma_{\mathrm{8.5~GHz}} = 3.8$~$\mu$Jy\,beam$^{-1}$. The synthesised beams are shown in the bottom-left corner of each map and have dimensions of $1.69 \times 1.57$~arcsec$^2$, $0.98 \times 0.90$~arcsec$^2$ and $0.53 \times 0.49$~arcsec$^2$. Greyscale: \textit{HST} NICMOS F160W image showing the two lensed images, the main lensing galaxy (edge-on spiral) and a nearby perturbing galaxy at the same redshift. Also shown are the critical curves (red) for the \citet{koopmans98} model that includes naked cusps.}
\label{fig:stuffr}
\end{center}
\end{figure*}

Maps were made from the combined and averaged data using \textsc{imagr}. The data were naturally weighted for maximum sensitivity and deconvolved using the CLEAN algorithm. The excellent sensitivity of the combined data leads to the detection of other sources within the telescope beam, particularly at 1.4~GHz where we cleaned 10 sources other than the lens. The residual continuum maps are shown in Fig.~\ref{fig:stuffr} for the 1.4-, 5- and 8.5-GHz data. The 8.5-GHz map is made up of data from 343 epochs and has a sensitivity of 3.8~$\mu$Jy\,beam$^{-1}$.

The 5 and 8.5~GHz maps both show an extended area of low-surface-brightness emission with a major axis that has a length and orientation similar to the separation vector of the two lensed images A and B, but which is offset by $\sim$0.5--1.0~arcsec to the north-east. The same structure is visible at 1.4~GHz, but here there is additional emission to the south and east. The subtraction of the lensed images has not been perfect as demonstrated by the sometimes prominent negative contours, but the similarity of the structures at the three separate frequencies makes us confident that the extended emission is real.

We have estimated the spectral index of the newly detected emission from the peak surface brightness. At each frequency this is located at approximately the same coordinate (the maximum separation is one 1.4-GHz pixel) and gives $\alpha = -1.0 \pm 0.1$ where $S_{\nu} \propto \nu^{\alpha}$. The emission is therefore very likely to arise from a faint jet that is associated with the radio core that forms the lensed images A and B, and at 8.5~GHz, the emission in fact looks like a jet emerging from image~A. If the lensed source is located very close to the radial caustic as advocated by \citet{koopmans98}, it is possible that the jet quickly leaves the multiply imaged region and is seen as unlensed emission close to image~A only, with no counterpart visible in the vicinity of image~B.

\subsection{Imaging of archival VLBI data}

\begin{figure*}
\begin{center}
\includegraphics[scale=0.5]{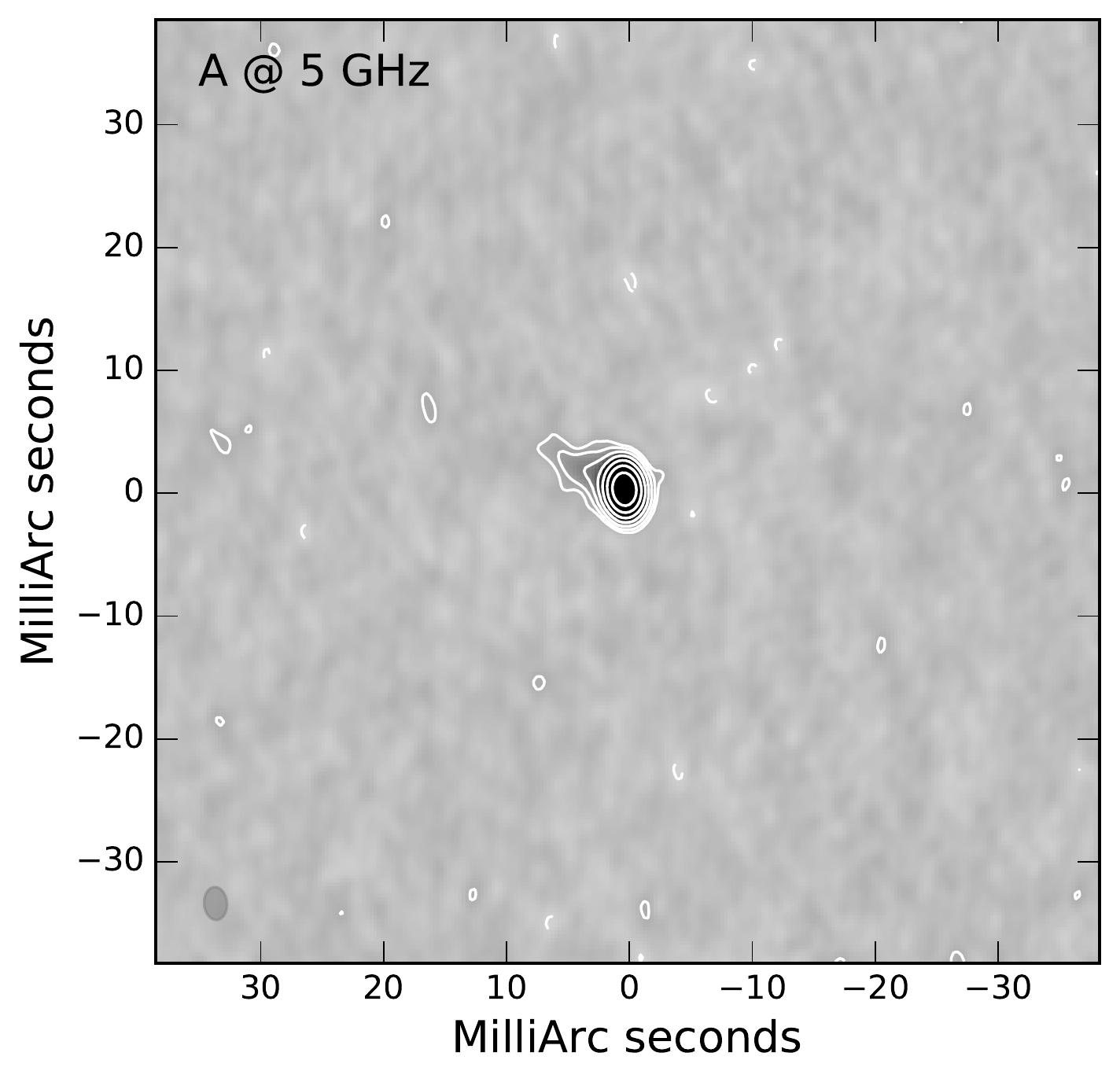}
\includegraphics[scale=0.5]{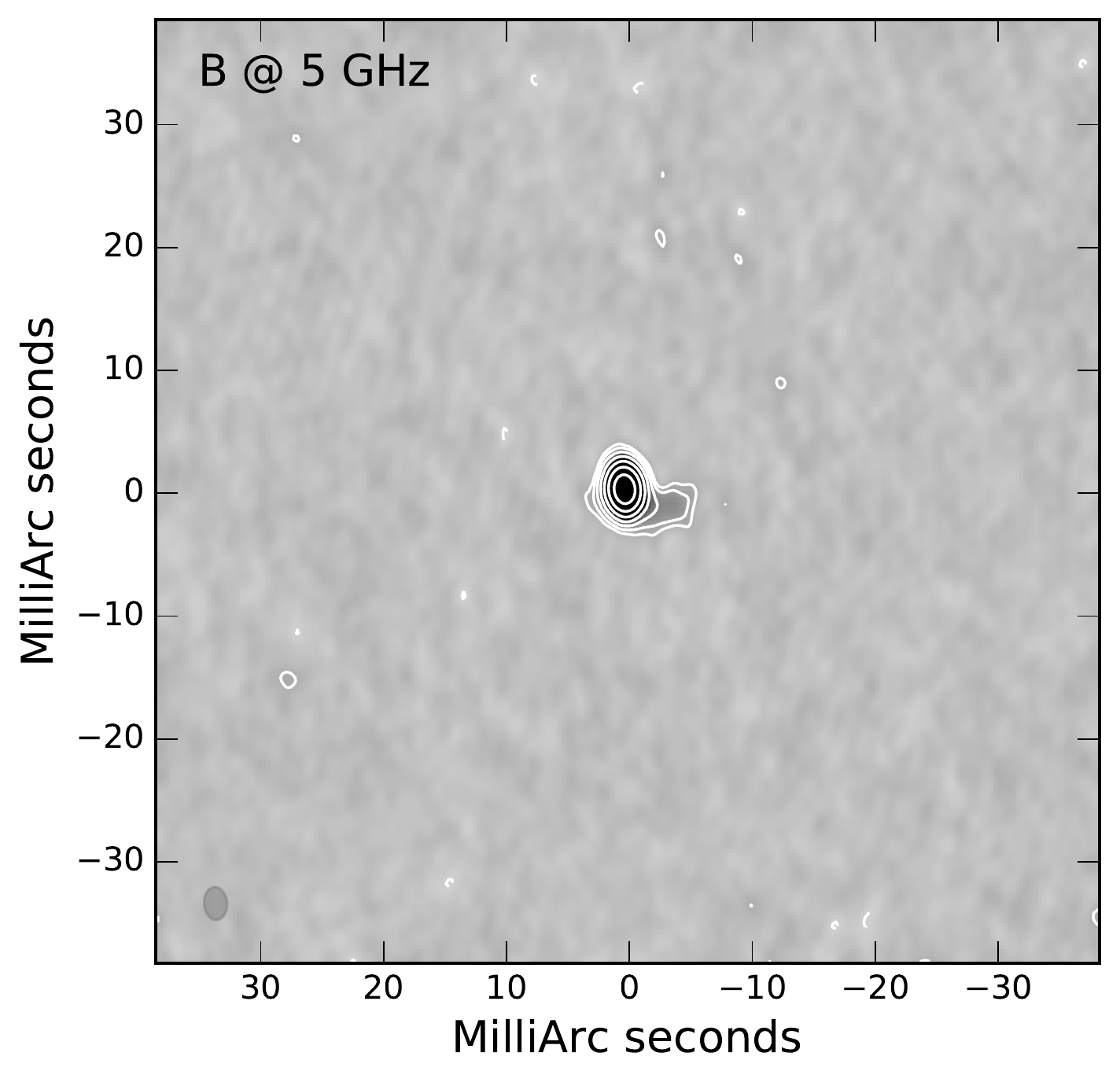}
\includegraphics[scale=0.5]{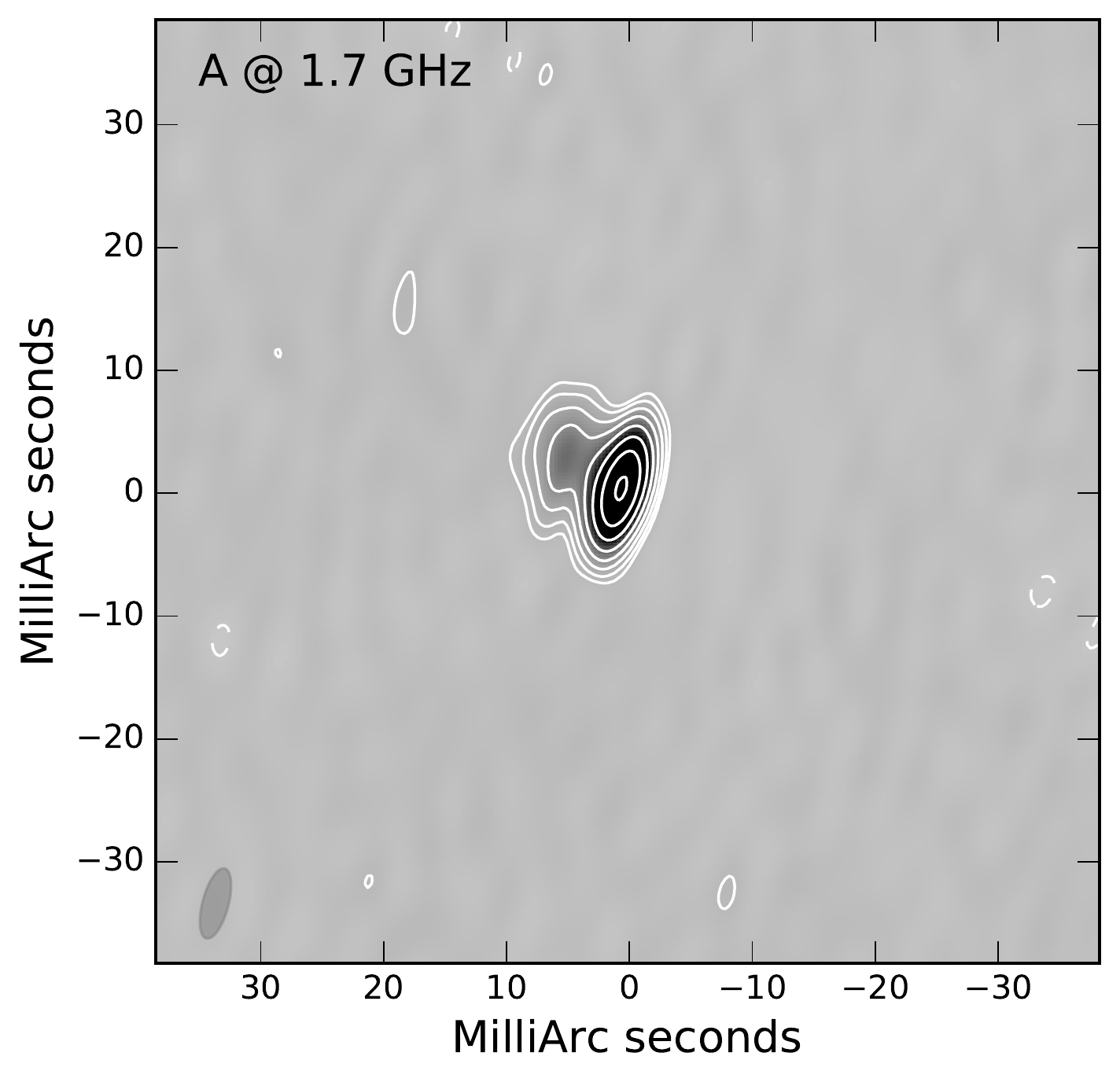}
\includegraphics[scale=0.5]{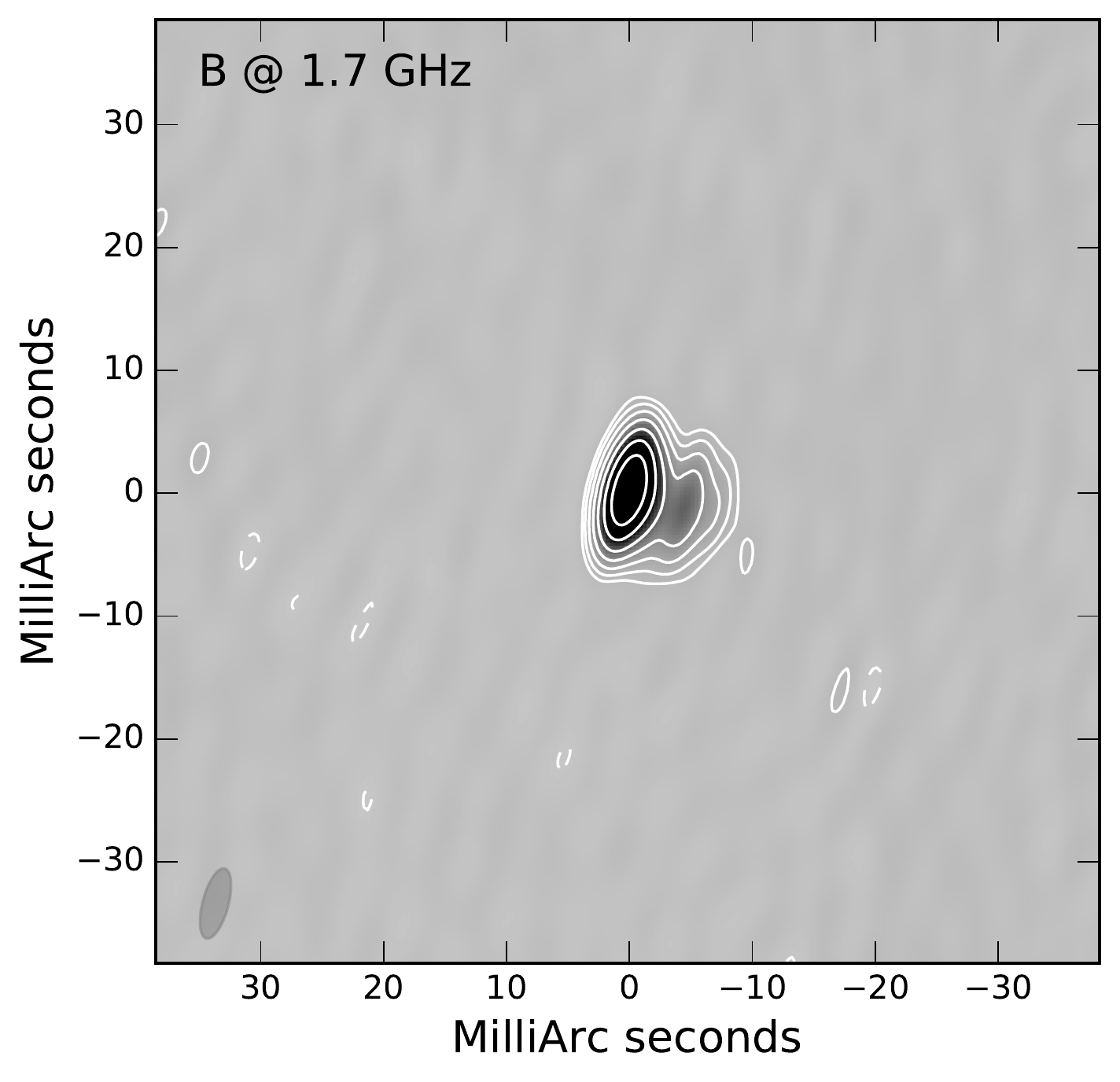}
\caption{VLBI maps of images A (left) and B (right). Top: VLBA 5-GHz maps ($\sigma = 82$~$\mu$Jy\,beam$^{-1}$, synthesized beam = $2.7 \times 1.9$~mas$^2$). Bottom: Global VLBI 1.7-GHz maps ($\sigma = 32$~$\mu$Jy\,beam$^{-1}$, synthesized beam = $5.9 \times 2.1$~mas$^2$). In all cases contours are plotted at $-1$, 1, 2, 4, etc. times the 3-$\sigma$ rms noise. At both frequencies short jets are seen in each image which point in nearly opposite directions.}
\label{fig:vlbi}
\end{center}
\end{figure*}

If the faint, extended emission is formed from a jet then this might be visible in VLBI maps. The only VLBI observations of B1600+434 published to date are the 15-GHz maps of \citet{patnaik01c} which show no sign of a jet on milliarcsec scales that could be the origin of the newly detected emission. However, any jet would be relatively faint at this frequency and thus we have produced our own maps from lower frequency archival data. One observation (Project code: BJ022) was made in February 1996 using the ten 25-m antennas of the Very Large Baseline Array (VLBA) at 5~GHz, and the other (Project code: GK020) at 1.7~GHz in February 2000 using a global VLBI array consisting of the VLBA together with six antennas of the European VLBI Network (EVN), including the 100-m Effelsberg and 76-m Lovell telescopes plus the Westerbork Synthesis Radio Telescope (WSRT) operating as a phased array. These data were calibrated and imaged in \textsc{aips} following standard procedures.

The resulting maps are shown in Fig.~\ref{fig:vlbi} and show a clear detection of a short ($\le$10~mas) jet at both frequencies and epochs. The jet in image~A points towards the northernmost part of the extended radio emission which supports the theory that this is jet emission associated with the lensed quasar. The direction of the jet as seen in each image also provides constraints on the lens model. In addition, if compact knots in the jet could be detected and their positions measured with milliarcsec precision, this would also allow the slope of the radial mass profile to be measured, a crucial parameter when measuring $H_0$ \citep{wucknitz01,suyu12}.

The VLBI data can also be used to explore why image~B appears to be entirely free of external variability. \citet{koopmans00b} postulate that the lower magnitude of variability is due to a lower average mass of the compact objects along the line of sight to B. This lies close to the stellar disc and bulge of the main lensing galaxy whereas image~A is seen predominantly through the galactic halo. Alternatively the microlensing might be suppressed by scatter-broadening in the ionised interstellar medium (ISM) of the lensing galaxy, or a combination of both.

The defining characteristic of scatter-broadening in an ionised ISM is a source size that increases with decreasing frequency as $\nu^{-2.2}$ \citep[e.g.][]{walker98}. The sizes of the VLBI cores were determined by modelfitting to the $u,v$ data discussed in Section~\ref{sec:lensmodel} using \textsc{difmap}. As this averages the data in frequency it is subject to bandwidth smearing and so we first formed a separate dataset for each image in \textsc{aips} by subtracting the clean components corresponding to the other image and then shifting the remaining image to the phase centre using \textsc{uvfix}. Once in \textsc{difmap}, a Gaussian component was used to represent the core and jet emission was removed using clean components between successive optimization of the parameters of the Gaussian.

\begin{table}
  \centering
  \caption{Source sizes ($\theta$) of Gaussian components fitted to the VLBI visibility data of images A and B at both 1.4 and 5~GHz. There is no evidence for a $\lambda^{2.2}$ dependence of the source size for either image.}
  \label{tab:vlbitheta}
  \begin{tabular}{cccc} \\ \hline
    Frequency (GHz) & $\theta_{\mathrm{A}}$ (mas$^2$) & $\theta_{\mathrm{B}}$ (mas$^2$) & $\theta_{\mathrm{A}}$ / $\theta_{\mathrm{B}}$ \\ \hline
    1.4 & $1.67 \times 1.21$ & $1.67 \times 1.19$ & 1.02 \\
    5   & $0.56 \times 0.36$ & $0.58 \times 0.39$ & 0.89 \\ \hline
  \end{tabular}
\end{table}

The results of the modelfitting are shown in Table~\ref{tab:vlbitheta}. Each image is resolved at both frequencies, a conclusion supported by Gaussian fitting in the image plane using \textsc{jmfit}. The sizes ($\theta$) of each image are very similar at 1.4~GHz with B being 12~per~cent larger than A at 5~GHz. B should be smaller than A by a factor of about 1.2 at both frequencies due to the lensing magnification, but we assume that this discrepancy is due to measurement error as if this was due to scatter-broadening, the effect should be larger at the lower frequency. In fact, the size of each image increases by approximately the ratio of the frequencies, a far weaker dependence than expected for scatter-broadening and thus we conclude that there is little evidence for this process occurring in B1600+434.

\subsection{A possible naked cusp}

As the VLBI data confirm that there is a jet pointing towards the faint, extended emission, it appears that after initially proceeding in a north-easterly direction the jet bends sharply by approximately 90\degr\ to continue towards the south-east. Even if this interpretation is true and we are seeing a predominantly non-lensed jet emanating from image~A, its sudden change in direction causes it to pass close to the centre of the lensing galaxy as seen in projection and this leaves the possibility that some of the jet emission is lensed, perhaps by a so-called `naked cusp' \citep[e.g.][]{kassiola93,evans98,rusin01}.

B1600+434 is a natural candidate for the presence of a naked cusp due to the main lensing galaxy being an edge-on spiral and thus forming a highly elliptical mass density distribution projected along the line of sight to the background quasar. For such flattened mass distributions, the inner (tangential) caustic can extend outside of the outer (radial) caustic, forming a pair of naked cusps on either side of the lens centre along the major axis of the lensing mass. Source emission which lies behind the naked cusp forms a characteristic configuration of three images on one side of the lens centre. If the source emission is extended, as is the case for a radio jet, the three images can merge to form an arc.

In Fig.~\ref{fig:stuffr} we also show the \textit{HST} NICMOS F160W image in which the edge-on spiral galaxy is very prominent\footnote{This image was taken as part of Proposal 9375 (PI C. Kochanek) and was observed on 2002 November 2. We downloaded the individual exposures from the \textit{HST} Data Archive and combined them into a final image using \textsc{AstroDrizzle}.}. This clearly demonstrates that the radio emission is located on one side of the lensing galaxy only, peaks along the major axis of the main lensing galaxy (especially noticeable at 1.4 and 5~GHz) and is extended in a direction orthogonal to the galaxy major axis -- this is the expected lensing configuration for a naked cusp. Fig.~\ref{fig:stuffr} also includes the critical curves of the \citet{koopmans98} lens model that includes naked cusps. Here it can be seen that the outer critical curve extends over the extended radio emission which, if the model is correct, strongly suggests that the faint radio emission is at least partially lensed.

Few examples of naked-cusp lensing have been observed. A prominent candidate is the luminous broad-absorption-line quasar APM~08279+5255 \citep{lewis02} where three approximately colinear images of the quasar are seen \citep{ibata99,egami00,lewis02a}, although no lensing galaxy has ever been detected \citep[e.g.][]{oya13}. Galaxy clusters should be more efficient at producing naked-cusp configurations \citep[e.g.][]{oguri04} and this seems to be the case with the lensed quasar SDSS~J1029+2623 \citep{oguri08}. B1600+434 may therefore be the first example of a naked-cusp lensing configuration in a galaxy-scale lens where the lensing geometry is clearly understood due to the position of the lensing galaxy being known. If the extended radio emission can be confirmed as being lensed, then additional lens constraints might become available.

\section{Conclusions}
\label{sec:conclusions}

We have presented an analysis of five seasons of VLA monitoring of the gravitational lens system CLASS~B1600+434 at 1.4, 5, 8.5 and 15~GHz, including polarization measurements at 8.5~GHz. The 1.4-GHz are not as suitable for time-delay analysis due to the relatively small magnitude of the variability and fewer epochs, but for the other data we have measured the time delay and its associated uncertainty using two different techniques. All results are consistent with one another, indicating that all sources of random and systematic error have been properly accounted for when simulating the variability curves used in our error analysis. Crucial in this regard has been the inclusion of realistic external variability into the radio light curves of image~A. Taking the measurement with the smallest uncertainty as our best result, we find a time delay of $42.3^{+2.0}_{-1.8}$ (random) $\pm 0.5$ (systematic)~d which, combining the random and systematic errors in quadrature, gives a final delay estimate of $42.3 \pm 2.1$~d (1-$\sigma$ uncertainty). Combining all delay estimates found using the CSM method gives a slightly higher value of $43.6\pm1.2$~d although we caution that the individual measurements may have correlated systematic errors.

Polarization monitoring has again proved able to measure a time delay in a gravitational lens system, there now being three systems for which this has been possible. The great benefit of polarization monitoring is the greater magnitude and shorter timescale of variability which is able to compensate for the large reduction in SNR compared to total flux density. The very smooth nature of the intrinsic total flux density variability in this source makes measuring the time delay difficult and future polarization monitoring with the current generation of broadband interferometer arrays might allow a significant improvement in the time-delay measurement for this system. We also find, however, that the polarized flux density is particularly affected by external variability and thus EVPA monitoring may hold the best prospects for improvement.

Unfortunately, it is far from clear that this lens system will ever yield a competitive measurement of $H_0$ due to the rather complicated lensing-mass distribution and the fact that two images provide very few constraints on the lens model. The long-term drift in the flux density ratio also has the consequence that this parameter's intrinsic value is not known with much certainty. We have though presented additional observations that might enable the construction of a more robust lens model than has been possible to date.

Firstly, we have presented the first maps that show the lensed VLBI jets in each image. Their orientations can be used as constraints on the lens model and if the relative lengths of the jets could be determined from the positions of subcomponents in the jet, this could allow a measurement of the slope of the radial mass profile in the lensing galaxy, an important parameter for $H_0$ determination. Secondly, we have detected previously unknown, very faint arcsec-scale structure in this system by combining all the epochs together to make extremely sensitive maps. The nature of this emission is not clear but it appears to be jet emission and its location suggests that it is related to the VLBI jet seen in image~A.

Whilst it is possible that the extended jet emission is simply the unlensed jet seen after it has left the multiple-imaging region, its location on one side of and close to the peak of the main lensing galaxy holds out the possibility that all or some of it is lensed and, in particular, formed by a so-called `naked cusp' configuration. B1600+434 has long been a likely candidate for a naked cusp due to the main lensing galaxy, an edge-on spiral, forming a very elongated projected mass profile. We recommend that this structure be imaged with higher angular resolution in order to better explore the lensing hypothesis, although the low surface brightness will make such an observation challenging. The e-MERLIN array operating at 1.4~GHz may provide the correct combination of sensitivity and angular resolution.

Finally, an examination of the modulation index of the external variability at the four frequencies supports the hypothesis originally put forward by \citet{koopmans00b} that the origin of this effect is microlensing of superluminal jet components by compact objects in the lensing galaxy. As well as the previously known external variability with a time-scale of days to weeks, we also find a longer-term change in the A/B flux density ratio on a time-scale of years, the simplest explanation for which is that it is also caused by microlensing. We find no evidence of external variability in image~B which will put stronger constraints on the mass function of the compact objects in the stellar disk and bulge.

\section*{Acknowledgements}

The author thanks Ian Browne for his many contributions to this project and the anonymous referee for a number of suggestions which improved the manuscript. The National Radio Astronomy Observatory is a facility of the National Science Foundation operated under cooperative agreement by Associated Universities, Inc. Based on observations made with the NASA/ESA \textit{Hubble Space Telescope}, obtained from the data archive at the Space Telescope Science Institute. STScI is operated by the Association of Universities for Research in Astronomy, Inc. under NASA contract NAS 5-26555. The European VLBI Network is a joint facility of independent European, African, Asian, and North American radio astronomy institutes. Scientific results from data presented in this publication are derived from the following EVN project code: GK020. The lens-model critical curves were digitized from the original postscript image using Version~4.4 of Ankit Rohatgi's \textsc{WebPlotDigitizer} software.

\section*{Data Availability}

The data underlying this article will be shared on reasonable request to the corresponding author.



\bibliographystyle{mnras}
\bibliography{lensing}


\bsp	
\label{lastpage}
\end{document}